\documentclass[a4paper,11pt]{article}
\usepackage{amsmath,amssymb,bm,mathrsfs,rotating,feynmp,pstricks,mathtools,cancel}
\usepackage{caption,subcaption}
\usepackage{color}
\pdfoutput=1 

\usepackage{jheppub} 

\usepackage[T1]{fontenc} 

\newcommand{\kahler}{K\"ahler\;}

\newcommand{\beq}{\begin{equation}}
\newcommand{\eeq}{\end{equation}}
\newcommand{\no}{\nonumber}
\renewcommand{\subsubsection}[1]{\addtocounter{subsubsection}{1}
\par\nobreak
\medskip
\nobreak
\noindent{\it \thesubsubsection.  #1 }
\par\nobreak\medskip\nobreak}

\def\lpar#1#2#3#4{\rlap{\raise#3\hbox{$\hskip#4#1\left\{\mbox{\phantom{\rule[0mm]{0mm}{#2}}}\right.$}}}
\def\rpar#1#2#3#4{\rlap{\raise#3\hbox{$\hskip#4\left\}#1\mbox{\phantom{\rule[0mm]{0mm}{#2}}}\right.$}}}

\title{\boldmath Models of Dynamical R-Parity Violation}

\author[a]{Csaba Cs\'aki,}
\affiliation[a]{Department of Physics, LEPP, Cornell University, Ithaca, NY 14853, USA}

\author[a]{Eric Kuflik,}

\author[b]{Oren Slone,}
\affiliation[b]{Raymond and Beverly Sackler School of Physics and Astronomy, Tel-Aviv University, Tel-Aviv 69978, Israel}

\author[b]{Tomer Volansky}

\abstract{The presence of R-parity violating interactions may relieve the tension between existing LHC constraints and natural supersymmetry. In this paper we lay down the theoretical framework and explore models of dynamical R-parity violation in which the breaking of R-parity is communicated to the visible sector by heavy messenger fields. We find that R-parity violation is often dominated by non-holomorphic operators that have so far been largely ignored, and might require a modification of the existing searches at the LHC. The dynamical origin implies that the effects of such operators are suppressed by the ratio of either  the light fermion masses or the supersymmetry breaking scale to the mediation scale, thereby providing a natural explanation for the smallness of R-parity violation. We consider various scenarios, classified by whether R-parity violation, flavor breaking and/or supersymmetry breaking are mediated by the same messenger fields. The most compact case, corresponding to a deformation of the so called flavor mediation scenario, allows for the mediation of supersymmetry breaking, R-parity breaking, and flavor symmetry breaking in a unified manner.
}

\begin{document}

\maketitle
\flushbottom

\section{Introduction\label{sec:Intro}}
For over two decades supersymmetry (SUSY) has been widely considered  the most likely extension of the Standard Model (SM) to be discovered once energy scales of a TeV are reached. Following the lack of an unambiguous supersymmetric signal at Run-I of the LHC~\cite{Aad:2014wea,Chatrchyan:2013iqa}, SUSY has lost some of its luster. The simplest SUSY model now requires sub-percent tuning in order to keep the Higgs mass stable at 125 GeV. However, many of the strong constraints apply strictly to the minimal extension, and there are relatively simple non-minimal models that significantly weaken or altogether eliminate the existing experimental constraints. One of the oldest and most commonly studied non-minimal framework is that of the Minimal Supersymmetric Standard Model (MSSM) supplemented  with R-parity violating (RPV) interactions~\cite{ Aulakh:1982yn,Hall:1983id,Ross:1984yg,Barger:1989rk,Dreiner:1997uz,Bhattacharyya:1997vv,Barbier:2004ez}. 

The original motivation for R-parity conservation is to forbid the presence of baryon (B) and lepton (L) number violating operators.  Naively, the most relevant of these interactions are the renormalizable holomorphic superpotential operators,
\beq
W_{\rm RPV} =  \mu_i l_i h_u + \lambda_{ijk}  \ell_i\ell_j\bar{e}_k +\lambda^\prime_{ijk}  \ell_i q_j \bar{d}_k+\lambda^{\prime\prime}_{ijk} \bar{u}_i\bar{d}_j\bar{d}_k \,.
\label{eq:WRPV}
\eeq
If the dimensionless coefficients $\lambda, \lambda', \lambda''$ are ${\cal O}(1)$, the operators above induce large baryon and lepton number violating processes, including proton decay, dinucleon decay and $n-\bar{n}$ oscillation~\cite{Dreiner:1997uz, Bhattacharyya:1997vv,Barbier:2004ez}. Thus, in the canonical MSSM, the operators (\ref{eq:WRPV})  are assumed to be completely absent due to an exact  $Z_2$ discrete symmetry, called R-parity, under which all SM fields are even and all superpartners are odd.
The realization of R-parity leads to the stability of the lightest supersymmetric particle (LSP) and implies that superpartners must be produced in pairs.

Surely, exact R-parity conservation is not required.   The constraints only imply that { R-parity is an approximate symmetry of the visible sector}, but it may be strongly broken elsewhere. 
By postulating that the couplings (\ref{eq:WRPV}) exist but are small, the approximate symmetry is apparent.   
However, this traditional framework of R-parity violation (RPV) is quite ad hoc, as the origin for the small couplings is unexplained. In addition, to ensure that RPV is phenomenologically relevant at colliders,  large hierarchies among the couplings must be invoked (for related models and phenomenological studies, see e.g., ~\cite{Nikolidakis:2007fc,Csaki:2011ge, Barbier:2004ez, Krnjaic:2012aj, Franceschini:2013ne, Csaki:2013we, Krnjaic:2013eta, Brust:2012uf, Graham:2012th, FileviezPerez:2012mj, Han:2012cu, Berger:2012mm, Franceschini:2012za, Ruderman:2012jd,Barger:2008wn,Masiero:1990uj} and references therein).   
A more attractive possibility is for R-parity to be exactly conserved at some high scale, while its breaking is communicated to the visible sector at a mediation scale, $M$.  Then R-parity is automatically an approximate symmetry, since  in the limit $M\rightarrow\infty$, the symmetry becomes exact in the visible sector.  The situation in which the violation of R-parity in the visible sector is dynamically generated at low-energies is termed {\em Dynamical R-Parity Violation} (dRPV)~\cite{Csaki:2013jza}. 

If RPV is dynamically generated, the superpotential terms of Eq.~\eqref{eq:WRPV}  are not expected to necessarily dominate the low energy theory.  Instead, much like the soft superpartner  masses, non-holomorphic interactions will also be  generated. 
As was shown in~\cite{Csaki:2013jza}, in many circumstances the dominant RPV operators appear in the \kahler potential and take the form
\beq 
\label{eq:KRPV}
{\cal O}_{\rm nhRPV} =  \eta_{ijk} \bar{u}_i\bar{e}_j\bar{d}_k^* + \eta^\prime_{ijk} q_i\bar{u}_j \ell_k^* +\eta^{\prime\prime}_{ijk} q_i q_j \bar{d}_k^*\,. 
\eeq
We will refer to the above operators as the "non-holomorphic dRPV operators". The effects of these operators on the low energy physics is automatically suppressed by a high mediation scale, as well as either  chirality or supersymmetry breaking, thereby providing a  leading order explanation for why the overall effects of RPV are small. 

The goal of this paper is to lay down the theoretical framework for building models of dRPV.    
 R-parity is taken to  be a good symmetry at some scale in order to forbid the unsuppressed holomorphic RPV superpotential terms of Eq.~(\ref{eq:WRPV}). The symmetry is then spontaneously broken by a field $S$.\footnote{For models of spontaneous R-parity violation in the visible sector see e.g.~\cite{SpontRPV}.} Heavy fields which couple to $S$  mediate the breaking to the visible sector, resulting in R-parity violating interactions. The  charges of $S$ under the global $U(1)_{B-L}$ and $U(1)_R$
 determine the form of the leading dRPV operators.   As we show, in some cases the holomorphic terms are allowed, but in most  cases the non-holomorphic dRPV operators dominate.
  Effects of supersymmetry breaking and flavor violation play a key role in the phenomenology of dRPV.   Supersymmetry may or may not be broken by the same dynamics responsible for the breaking of R-parity, and correspondingly the breaking may or may not be communicated by the same messenger fields.   Similarly, the flavor structure of dRPV operators may be  affected by any  dynamics that breaks the approximate $U(3)^5$ flavor symmetry of the SM and can provide an explanation for the origin of its flavor structure.  As with supersymmetry breaking, a dynamical origin of flavor breaking may also be communicated to the visible sector together with RPV.  The realization and implications of  the above possibilities will be discussed below. 
  
The paper is organized as follows.  In {\bf Sec.~\ref{sec:BasicdRPV}} we present some of the basic properties of dRPV models. We show that the effects of the non-holomoprhic dRPV operators are either chirally suppressed, or suppressed by SUSY breaking. We also explain how the low energy $U(1)_{B-L}$ and $U(1)_R$ symmetries can be used to determine whether holomorphic  or non-holomorphic dRPV will be dominant in the effective theory. 
{\bf Sec.~\ref{sec:GeneratingdRPV}} is devoted to presenting several toy model examples of dRPV, which exemplify the key features of the model-building. In some of these models, R-parity is broken in a secluded hidden sector, while in others there is a direct coupling between the SM fields and the field responsible for the breaking of R-parity.   
We end this section by discussing how to extend the toy models to a complete one.
   In {\bf Sec.~\ref{sec:SUSYbreaking}} we analyze the consequences of SUSY breaking on the dRPV operators.  We distinguish between two cases: (i) The RPV sector is supersymmetric and SUSY-breaking is external to it,  and (ii) the RPV sector is directly coupled to the SUSY-breaking sector. 
 {\bf Sec.~\ref{sec:flavordrpv}} contains our analysis of the flavor structure  of the dRPV terms. As an illustration, we assume that the SM flavor hierarchy is generated via the Froggatt-Nielsen mechanism~\cite{Froggatt:1978nt} in which a horizontal flavor symmetry is spontaneously broken. 
We consider the cases where the dynamics which breaks and communicates the flavor symmetry is either secluded from or unified with the R-parity breaking sector.  
In {\bf Sec.~\ref{sec:FlavorMediation}} we study an extension of a  model of flavor mediated supersymmetry breaking~\cite{Kaplan:1998jk, Kaplan:1999iq, Shadmi:2011hs, Abdullah:2012tq, Galon:2013jba, Nomura:2008gg, Nomura:2008pt, Craig:2012di, Pomarol:1995xc, Grossman:2007bd}, which allows for  the unification of SUSY, R-parity and flavor symmetry breaking sectors. 
We conclude in {\bf Sec.~\ref{sec:Conclusions}}. In {\bf Appendix~\ref{sec:FullModel}} we present some of the details of a complete UV model, and the constraints arising from proton decay, neutrino masses and the LSP lifetime.  {\bf Appendix~\ref{App:FNsuppression}} contains the tables of flavor suppressions for a particularly successful Froggatt-Nielsen model, and {\bf Appendix~\ref{App:SUSYidentities}} contains some useful supersymmetric identities used throughout the paper.

\section{Basic Properties of dRPV\label{sec:BasicdRPV}}

As mentioned above, the dRPV paradigm  provides a mechanism which explains the smallness of R-parity violating  effects and often implies the dominance of the operators of Eq.~\eqref{eq:KRPV}.  Before introducing models of dynamical RPV, we make some general comments on the low energy effective theory that clarify these statements.

\subsection{Suppression of RPV Effects}
\label{sec:Suppression}

The interactions of the dRPV operators (\ref{eq:KRPV}) are either chirally suppressed or suppressed by SUSY-breaking, which can be easily seen by using the equations of motion. To illustrate this we consider the $qq\bar d^*$ operator.    Since in the dRPV framework R-parity is broken spontaneously by a field, $S$, and the breaking is mediated to the visible sector at a scale $M$, such an operator is proportional to $S / M$.       
  First, the dRPV operator, which resides in the \kahler potential, is expressed as an F-term according to (\ref{eq:DtoF}). Then, using the equation of motion (\ref{eq:EOM}) for $\bar{d}$ one finds,
\beq
- \int d^4\theta\,  \frac{S^*}{|M|^2} q q \bar{d}^* =  \int d^2\theta\,  \frac{ qq }{|M|^2}   \frac{1}{4}\mathcal{D}^{\dagger 2} (S^*\bar{d}^*) \,\mathop{\longrightarrow}^{EOM} \int d^2\theta \left[ \frac{  \langle S \rangle^*}{|M|^2}  q q \frac{\partial W }{\partial \bar{d}} +  \frac{  \langle F_S \rangle^*}{|M|^2}  q q \bar{d}^* \right]\,.\label{eomequiv}
\eeq
The first term above induces chirally suppressed operators, while the second term is responsible for SUSY-breaking suppressed operators assuming $S$ has a non-vanishing $F$-term VEV. 
We will see in Sec.~\ref{sec:SUSYbreaking} that the contribution similar to the last term above  will appear in the absence of $F_S$, when supersymmetry breaking is secluded from the dynamics inducing the spontaneous breaking of R-parity.  
 Similar considerations show that if the standard holomorphic operators of Eq.~\eqref{eq:WRPV} originate  from a K\"ahler term, they too will be suppressed by either chiral symmetry or by supersymmetry  breaking.

The above discussion suggests that in the limit of unbroken supersymmetry and vanishing Yukawa couplings, the dRPV operators may not  induce R-parity violating interactions. To see this, we can  consider the simple theory with canonical kinetic terms for $q$ and $d$, and the above dRPV operator,
\beq
K = |\bar{d}|^2 + |q|^2 + \frac{S^*}{|M|^2} q q \bar{d}^* +h.c.\,.
\eeq 
Under the non-linear field redefinition $\bar{d} \rightarrow \bar{d} - \langle S^* \rangle {q q }/{|M|^2}$, the K\"ahler potential above becomes
\beq
K = |\bar{d}|^2 + |q|^2 + \left(\frac{S^*-\langle S^* \rangle}{|M|^2}qq\bar{d}^* +h.c. \right)+ \left| \frac{\langle S \rangle}{M^2}\right|^2 |q|^4\,.
\eeq
In the absence of the standard Yukawa coupling, $h_d q \bar{d}$, or an $F$-term for $S$,  this transformation completely removes  R-parity violating interactions from the theory.   Conversely, for $F_S\neq 0$, the terms in the brackets induce RPV terms suppressed by the SUSY-breaking $F_S$ term, while in the presence of the Yukawa couplings, the above field redefinition results in the additional term $\langle S^* \rangle  h_d qqq /|M|^2 $ in the superpotential. This provides another simple argument that all RPV interactions originating from the dRPV operators must be either suppressed by SUSY-breaking or by the light fermion masses. This may also be verified by a brute-force component expansion of a the dRPV operators, as shown in (\ref{kahlerexpand}).

\subsection{Spurions and Symmetries\label{sec:spurion}}

To understand  whether non-holomorphic or holomorphic RPV operators will dominate at low energy, we must consider the symmetries of the low energy effective action. The MSSM without supersymmetry breaking has three 
global $U(1)$ symmetries in addition to hypercharge: baryon number, lepton number and a continuous $R$-symmetry. The SM matter fields $q,\bar{u},\bar{d},\ell,\bar{e}$ can be taken to have charge $1/2$ under the R-symmetry, while the Higgs fields $h_{u},h_d$ have R-charge 1.  The $B$ and $L$ charges are standard.  

In order to  identify  the  leading RPV operators,  one may utilize a spurion analysis in the low energy effective theory. A crucial observation is that the $U(1)_{B-L}$ and $U(1)_{R}$  symmetries  differentiate between the holomorphic, Eq.~\eqref{eq:WRPV}, and non-holomorphic,  Eq.~\eqref{eq:KRPV}, RPV operators:
\begin{equation}\begin{array}{ccc}
& {\rm U}(1)_{B-L}	& {\rm U}(1)_R\\
{\mathcal{O}_{nhRPV}}: & 1 & 1/2 \\
{\mathcal{O}_{hRPV}}: & -1 & 3/2 
\end{array}\end{equation}
Therefore, it is the $U(1)_{B-L}$ and $U(1)_{R}$ charge of the spurions responsible for their breaking that will determine which of the operators dominate at low energy.   
For instance,  if the RPV breaking field, $S$, has $U(1)_{B-L}$ and $U(1)_{R}$ charges $1$ and $1/2$ respectively, then both the dRPV K\"ahler potential operators $\frac{S^*}{M}  \mathcal{O}_{nhRPV}$ and the traditional superpotential operators $\frac{S}{M^2}  \mathcal{O}_{hRPV}$ may appear in the effective theory.  Since the effects of the dRPV operators are chirally or SUSY-breaking suppressed, these spurion charge assignments will result in effective theories where the leading source of RPV are the holomorphic terms. This will be the situation in the first toy model we present in Sec.~\ref{sec:toymodel}.  

If 
 $S$ carries $U(1)_{B-L}$ and $U(1)_{R}$ charges $-1$ and $-1/2$ respectively, then the non-holomorphic dRPV operator $\frac{S}{M^2}  \mathcal{O}_{nhRPV}$ will be allowed while no holomorphic operator will be generated,  leading to non-holomorphic dRPV domination at low energy. This will be the case in the model presented in Sec.~\ref{sec:goodmodel}.
In Sec.~\ref{sec:moremodels},  we will discuss two additional simple models.  In the first, $S$ carries $U(1)_{ B-L}$ and $U(1)_{ R}$ charges $1/2$ and $1/2$, leading to the dominant RPV term given by the dimension 6 superpotential operator $ S h_d q q q $.
In the second, $S$ is charged  $3/4$ and $1/2$ under the above symmetries, leading to the dominant RPV term given by the 
dimension-7 K\"ahler operator $ S^2 \mathcal{D}_\alpha^2 \mathcal{O}_{nhRPV}$.

\section{Generating Effective dRPV\label{sec:GeneratingdRPV}}

In this section we present the basic structure of dRPV models.   Most of the models studied should be thought of as  toy models  that capture the essence of  dynamically generating non-holomorphic RPV terms in the K\"ahler potential.   The theories are renormalizable and give rise to  non-renormalizable, non-holomorphic operators  in the low energy effective action once messenger fields are integrated out. 
Extending the toy models to more complete ones is straightforward and requires the introduction of flavor indices, which we suppress in the toy models, as well as supplementing the superpotentials with additional interactions that are responsible for the generation of dRPV operators not present in the simplified examples.   We discuss a complete model in the last subsection.

\subsection{A Toy Model\label{sec:toymodel}}

Consider a theory in which  R-parity is preserved in the UV and spontaneously broken by the VEV of the field $S$.  The low energy effective theory will contain R-parity violating operators.  As a first illustration, let us study the superpotential,
\beq
W = M D \bar{D}+ S  D  \bar{d} + \lambda q q D \label{toymodel} \ .
\eeq
 Here    $\bar D$ and $D$ are vector-like  fields which carry the SM quantum numbers of the $\bar d$ and its complex conjugate, respectively.  For now, we deliberately ignore the effects of supersymmetry breaking and assume a canonical \kahler potential.   Above, and throughout the paper, we denote heavy fields by upper-case letters, while lower-case letters denote the light MSSM degrees of freedom.

The fields $q, \bar{d}$ and $S$ are R-parity odd, while $D$ and $\bar{D}$ are R-parity even. Upon integration out of the heavy R-parity neutral fields, the entire effective superpotential is set to zero (which can be straightforwardly seen from the $D$ equation of motion (EOM)).   Meanwhile, the effective low energy \kahler potential is non-trivial and takes the form  
\begin{equation}
K_{eff}  =  |q|^2 + |\bar{d}|^2+ \frac{1}{|M|^2}  \left|S \bar{d} + \lambda qq  \right|^2 +   \mathcal{O} \left(\frac{1}{M^{4}}\right) \,,
  \label{toyeff}
\end{equation}
where the cross-term contains  the non-holomorphic dRPV coupling,
\beq
K_{dRPV} = \frac{\lambda S^*}{|M|^2} q q \bar{d}^* +h.c. 
\label{qqd}
\eeq
The above simple model dynamically induces R-parity violation, once the heavy fields are integrated out and $S$ obtains a VEV.

The origin of  R-parity violation in this model is the mixing among the heavy and the light fields; there is an  order $\langle S\rangle /M$ mixing between $\bar{d}$ and $\bar{D}$, and  a sub-leading mixing between the heavy $D$ and the light multiplet $\bar{d}^*$. The latter mixing can be seen by considering the sub-leading term of the $\bar D$  EOM,
$M D=-\frac{1}{4} {\mathcal{D}^{\dagger\, 2}_\alpha}\bar{D}^*$, which together with the leading $D$ EOM implies a small mixing between $D$ and $\bar{d}^*$ of the form,
\begin{equation}
D \sim -\frac{1}{4} \frac{\langle S \rangle^*}{|M|^2} \mathcal{D}^{\dagger 2} \bar{d}^*  \ .
\label{eq:unusualmixing}
\end{equation}
This atypical mixing is between the fermion of $D$ and a derivative of the fermion of $\bar{d}^*$, as well as  between the F-terms and derivatives of the scalars, (see Eq.~\ref{eq:Dsquare}). Due to the derivatives, one finds that~\eqref{eq:unusualmixing} is chirally suppressed.
Substituting the mixing above in the last term of Eq.~\eqref{toymodel}, the dRPV \kahler potential, Eq.~\eqref{qqd}, is restored,
\beq
\int d^2\theta\, \lambda q q D\longrightarrow \int d^2\theta\,\lambda q q \,\left(-  \frac{\langle S \rangle^*}{4|M|^2} \mathcal{D}^{\dagger 2} \bar{d}^* \right) = \int d^4\theta\,  \frac{\lambda \langle S \rangle^*}{|M|^2} 
q q \bar{d}^* \,.
\eeq 

\subsection{dRPV from a Hidden Sector}
\label{sec:goodmodel}

The toy model in the previous section has an immediate disadvantage; it provides no explanation as to why the non-holomorphic RPV is dominant at low energy.   For instance, the term $\bar{u}\bar{d}\bar{D}$ is invariant under  $U(1)_{B-L}$ and $U(1)_{R}$, and could have been added to the Lagrangian. In the presence of such a term, once $D$ and $\bar D$ are integrated out, an effective holomorphic RPV term is generated,
\begin{equation}
W_{eff} = -\frac{S}{M} \bar{u}\bar{d}\bar{d}\,.
\end{equation}
Since this term does not exhibit any further chiral suppressions,  the effect of this operator will generically dominate over the effect of the non-holomorphic operator.   Indeed, a vertex from the non-holomorphic term will be additionally suppressed by $m_d /M$ compared to the holomorphic vertex (where $m_d$ is the down-type fermion mass). The presence of a leading holomorphic RPV term is in accordance with our expectations from the symmetry considerations discussed in Section \ref{sec:spurion}: the charges of the RPV field $S$ are $1$ and $1/2$ under $U(1)_{B-L}$ and $U(1)_R$, and thus large holomorphic RPV terms are allowed.

In the above toy model, $S$ directly couples to the visible sector and its VEV induces a mixing between the heavy and light fields.   It is interesting to consider a deformation of this model,  in which the sector responsible for the spontaneous breaking of R-parity is secluded from the visible one and the heavy fields act as messengers which communicate the breaking. A dynamical hidden sector model with suppressed holomorphic couplings can be easily realized,  with a mild modification of the previous toy model.  
We introduce a second set of heavy messengers, $D_-, \bar{D}_-$, which carry negative R-parity charge, while the original (positively charged) messengers are denoted $D_+,\bar{D}_+$.   The RPV field $S$ triggers a mixing between the two types of messengers, but does not interact with the visible fields at the renormalizable level. The superpotential of the model is,
\begin{equation}
W=M(D_+ \bar{D}_+ +D_-\bar{D}_-) + S \bar{D}_+ D_- + m D_-\bar{d} + qq D_+\, . \label{improvedtoy}
\end{equation}
Integrating out the heavy messenger sector  does not generate an effective superpotential term at the leading order, and the leading RPV K\"ahler term is given by
\begin{equation}
\frac{\langle S\rangle  m^*}{|M|^2 M} qq \bar{d}^* +h.c. 
\label{eq:nonholohidden}
\end{equation}
The above is simple to understand.  The dRPV operator must be proportional to the VEV of the RPV-breaking field, and to $m$, since it controls the mixing which connects the heavy and light sectors. 

Just as in the original toy model of the previous section, the $U(1)_{B-L}$ and $U(1)_R$ symmetries allows for the additional superpotential term $\bar{u}\bar{d} \bar{D}_+$. The effective superpotential will no longer vanish in the presence of this term,
\begin{equation}
W_{eff}= \frac{1}{M} \bar{u}\bar{d} qq\,.
\end{equation}
However, this term does not violate R-parity. This is again in accordance with our general expectation from the spurion analysis: now the charges of $S$ are $-1$ and $-1/2$ under $U(1)_{B-L}$ and $U(1)_R$, which implies a suppression in the holomorphic RPV terms. By carefully taking into account the effect of the $\bar{u}\bar{d} \bar{D}_+$ operator on the K\"ahler potential, one finds that the leading holomorphic RPV operator is given by
\begin{equation}
\frac{m}{|M|^4} \langle S^* \rangle (\mathcal{D}_\alpha^2 \bar{d}) \bar{u}\bar{d}\,,
\end{equation}
which is indeed allowed by the symmetries, but is more strongly suppressed than the non-holomorphic term in (\ref{eq:nonholohidden}).

\subsection{More Toy Models}
\label{sec:moremodels}

Let us now briefly study additional variations of dRPV models. In the first example, $S$ carries charges $(1/2,1/2)$ under $U(1)_{B-L}\times U(1)_{R}$, leading to the dominant RPV term given by the dimension 6 superpotential operator $ \frac{1}{S} h_d q q q $. In the second model, $S$ has charges $(3/4,-1/2)$ which leads to the dominant RPV term given by  
dimension-7 K\"ahler operators of the form $ S^2 \mathcal{D}_\alpha^2 \mathcal{O}_{nhRPV}$. The resulting operators are suppressed by either two powers of light fermion masses, one power of fermion mass and supersymmetry breaking, or two powers of the supersymmetry breaking scale. As with the previous examples, both of the toy models below can be extended to include the full MSSM by following the steps described in Sec.~\ref{sec:complete}.

\subsubsection{Superpotential dRPV\label{sec:suppotdRPV}}

Consider the superpotential,   
\beq
W = S D \bar{D} + h_d q \bar{D} + q q \bar{D}.
\eeq
Above, $\bar{D}$ and $\bar{d}$ have the same quantum numbers and the model differs from the previous examples in that the  mixing between the heavy and light fields is induced by electroweak symmetry breaking.
 This model also has the virtue of being a hidden-sector model, as there is no direct coupling between $S$ and the MSSM fields. Integrating out the heavy fields leads to the low energy superpotential,
\beq
W_{eff} = \frac{1}{S} h_d q q q.
\eeq
As explained in Sec.~\ref{eomequiv}, the phenomenological effects of this operator are similar to those of the dRPV term $qq\bar{d}^*$ in the absence of supersymmetry breaking.

\subsubsection{Doubly Suppressed dRPV\label{sec:doublysuppressed}}

Here we consider a model analogous to the original toy model, but instead of having heavy $D$'s that mix with light fields, heavy doublet $Q$'s are introduced together with the superpotential,
\beq
 W =  M Q \bar{Q} +S \bar{Q} q + \lambda \bar{Q} \bar{Q} \bar{d} \ . 
 \label{doublesupw}
\eeq
The heavy field $\bar Q$ mixes with the light one via, 
\begin{equation}
\bar{Q} \sim -\frac{1}{4} \frac{ 1}{|M|^2} \mathcal{D}^{\dagger 2} S^* \bar{q}^*  \ .
\label{eq:unusualmixing2}
\end{equation}
Since  two mixings of this form
are needed in Eq.~\eqref{eq:unusualmixing2} in order to induce the low energy Lagrangian, the resulting dRPV couplings are doubly suppressed by chiral symmetry breaking (or  by the supersymmetry  breaking scale). 
Indeed integrating out the fields results in a vanishing superpotential, $W=0$, and  \kahler potential,
\beq
K= |\bar{d}|^2+|q|^2 \left(1+ \frac{|S|^2}{|M|^2} \right)- \frac{1}{4}  \frac{\lambda^* S }{|M|^4}  (\mathcal{D}^{ 2}S q)q \bar{d}^*+h.c. \,.
\eeq
The dRPV term can be simplified by using the EOM,
\beq
-\frac{1}{4} \frac{\lambda^* S}{|M|^4}  (\mathcal{D}^{ 2}S q)q \bar{d}^* \rightarrow
 \frac{\lambda^* \langle S\rangle   
 }{|M|^4} \left(\langle F_{S}\rangle  qq\bar{d}^* - \langle S\rangle  \frac{\partial W^*}{\partial q^*} q \bar{d}^* \right) .
\eeq
The first of the resulting two terms is the standard non-holomorphic dRPV term with additional supersymmetry breaking suppression, while the second term can lead to operators suppressed by two fermion masses or one fermion mass and supersymmetry breaking. Thus, the RPV couplings are hierarchical and highly  suppressed.  

The case where the supersymmetry breaking effects are subdominant, offers an interesting possibility where the 3rd generation RPV couplings are much larger than the other couplings. Using the EOMs, the second term can be written  as an F-term,
\beq
\int d^2\theta\frac{\lambda \langle S^* \rangle ^2}{|M|^2}\frac{\left\langle h_uh_d \right\rangle }{M^2}y_{d}y_u\bar{u} \bar{d} \bar{d} \,.
\eeq
One can see that the effect is equivalent to a traditional $\bar{u}\bar{d}\bar{d}$ RPV operator, except that it arises from a dimension-5 superpotential term with a very strong flavor suppression, similar to~\cite{Csaki:2011ge}. The main difference between this model and that of~\cite{Csaki:2011ge}  is that here there are two powers of the Yukawa suppressions instead of three powers, however the model at hand enjoys additional overall suppressions due to the mediation scale, $M$. Also, the large flavor hierarchy here  is independent of any assumptions about the flavor dynamics.

\subsection{A Complete Model}
\label{sec:complete}

The ingredients presented above can be incorporated into a complete model including both quarks and leptons.  To illustrate this, consider the toy example given in Eq.~\eqref{toymodel}.    A minimal extension which  includes the full set  of dRPV operators given in  Eq.~\eqref{eq:KRPV} requires  the introduction of at least one more vector-like pair of heavy messenger fields, $L$ and $\bar{L}$, which carry the SM quantum numbers of the lepton doublet $\ell$ and its complex conjugate, respectively. 
The superpotential that generates the full dPRV MSSM is
  \beq\begin{array}{rcl}
W &=& M_{D_i} D_i \bar{D}_i+ M_{L_i} L_i \bar{L}_i + \lambda^{d}_{ij} S  D_i  \bar{d}_j +  \lambda^{\ell}_{ij}  S \bar{L}_i \ell_j  \\
&&+ \gamma_{ijk} \bar{u}_i \bar{e}_j D_k +  \gamma_{ijk}^{\prime} q_i \bar{u}_j \bar{L}_k +  \frac{1}{2}\gamma_{ijk}^{\prime \prime}  q_i q_j D_k\\
&&+ W_{\rm MSSM}\ , \label{toyfull}
\end{array}\eeq 
where in general there can be an arbitrary number of heavy messengers.\footnote{Note that the heavy messengers furnish a ${\bf 5_H + \bar{5}_H}$ pair under an SU(5) GUT group. The superpotential of Eq.~\eqref{toyfull} can be summarized in the SU(5) language (assuming a single coupling) simply as 
\[
W= M {\bf 5_H \bar{5}_H} + S {\bf 5_H \bar{5}_L} +  {\bf 10_L 10_L 5_H} +W_{MSSM}\ .
\]
The other models presented above can be similarly generalized.}
Working with the low energy effective theory and including the flavor indices and effects of SUSY-breaking,  the superpotential~\eqref{toyfull} results in  the effective Lagrangian
\begin{eqnarray}{}
W_{\rm eff} &=& W_{\rm MSSM} \no\\
K_{\rm eff} & = & |q_i|^2 +|\bar{u}_i|^2+|\bar{e}_i|^2 + \left( \delta_{ij}+ \alpha_{ij}^{d} \frac{|S|^2}{M^2} \right) \bar{d}^{*}_j \bar{d}_i + \left( \delta_{ij}+ \alpha_{ij}^{\ell} \frac{|S|^2}{M^2} \right) \ell^{*}_j \ell_i\no\\
  &&+\eta_{ijk}  \frac{S^*}{M^2} \bar{u}_i\bar{e}_j\bar{d}^{*}_k + 
   \eta^\prime_{ijk} \frac{S^*}{M^2}  q_i\bar{u}_j \ell^{*}_k +
  \frac{1}{2}\eta^{\prime\prime}_{ijk}  \frac{S^*}{M^2} q_i q_j \bar{d}^{*}_k +h.c.\no\\
     &&+\frac{1 }{M^{2}}\left(\frac{1}{4}\lambda^{\prime\prime}_{ijmn}q_i q_j q_m^* q_n^*  +\lambda^{\prime}_{ijmn} q_i\bar{u}_j q_m^* \bar{u}_n^* + \lambda^{}_{ijmn}\bar{u}_i \bar{e}_j \bar{u}_m^* \bar{e}_n^*   \right. \no\\   
  &&~~~~~~~~~+ \left. \frac{1}{2} \lambda^{\not{B},\not{L}}_{ijmn}q_iq_j \bar{u}^*_m\bar{e}^*_n + h.c. \right)+\mathcal{O} (M^{-4})\,, 
  \label{Kdrpv2}
\end{eqnarray}
where the low energy coupling constants depend on the UV couplings of Eq.~\eqref{toyfull} and are given  by
\beq \arraycolsep=1.4pt\def\arraystretch{2.2} \begin{array}{ccc}
\alpha_{ij}^d = \dfrac{M^2}{M_{D_k}^2} \lambda_{ki}^d\lambda_{kj}^{d*},~~~~\alpha_{ij}^\ell = \dfrac{M^2}{M_{L_k}^2} \lambda_{ki}^\ell\lambda_{kj}^{\ell*}\,,\\
\eta_{ijk} = \dfrac{M^2}{M_{D_m}^2} \gamma_{ijm} \lambda_{mk}^{d*},~~~~\eta^{\prime}_{ijk} = \dfrac{M^2}{M_{L_m}^2} \gamma^{\prime}_{ijm} \lambda_{mk}^{\ell*},~~~~\eta^{\prime\prime}_{ijk} = \dfrac{M^2}{M_{D_m}^2} \gamma^{\prime\prime}_{ijm} \lambda_{mk}^{d*} \,,\\
{\lambda_{ijmn}^{\prime\prime}} = \dfrac{M^2}{M_{D_k}^2} \gamma_{ijk}^{\prime\prime} \gamma_{lmk}^{\prime\prime*},
~~~~{\lambda_{ijmn}} = \dfrac{ M^2}{M_{D_k}^2} \gamma_{ijk}^{} \gamma_{lmk}^{*}\,,\\
{\lambda_{ijmn}^{\not{B},\not{L}}} = \dfrac{M^2}{M_{D_k}^2} \gamma_{ijk}^{\prime\prime} \gamma_{lmk}^{*},
~~~~{\lambda_{ijmn}^{\prime}}=\dfrac{M^2}{M_{L_k}^2} \gamma_{ijk}^{\prime} \gamma_{lmk}^{\prime *}\,.
 \label{noflavordef}
\end{array}\eeq

As will be discussed in Sec.~\ref{sec:flavordrpv}, several generations of the messenger fields can be motivated if each carry different charges under a horizontal flavor symmetry and mediate the flavor structure to the SM. A detailed exploration of the model in Eq.~(\ref{toyfull}) is presented in App.~\ref{sec:FullModel} where we derive constraints arising from proton decay, neutrino masses and the LSP lifetime.  

\section{SUSY breaking Effects\label{sec:SUSYbreaking}}

\begin{figure}[t!]
\begin{center}
\includegraphics[width=.46\textwidth]{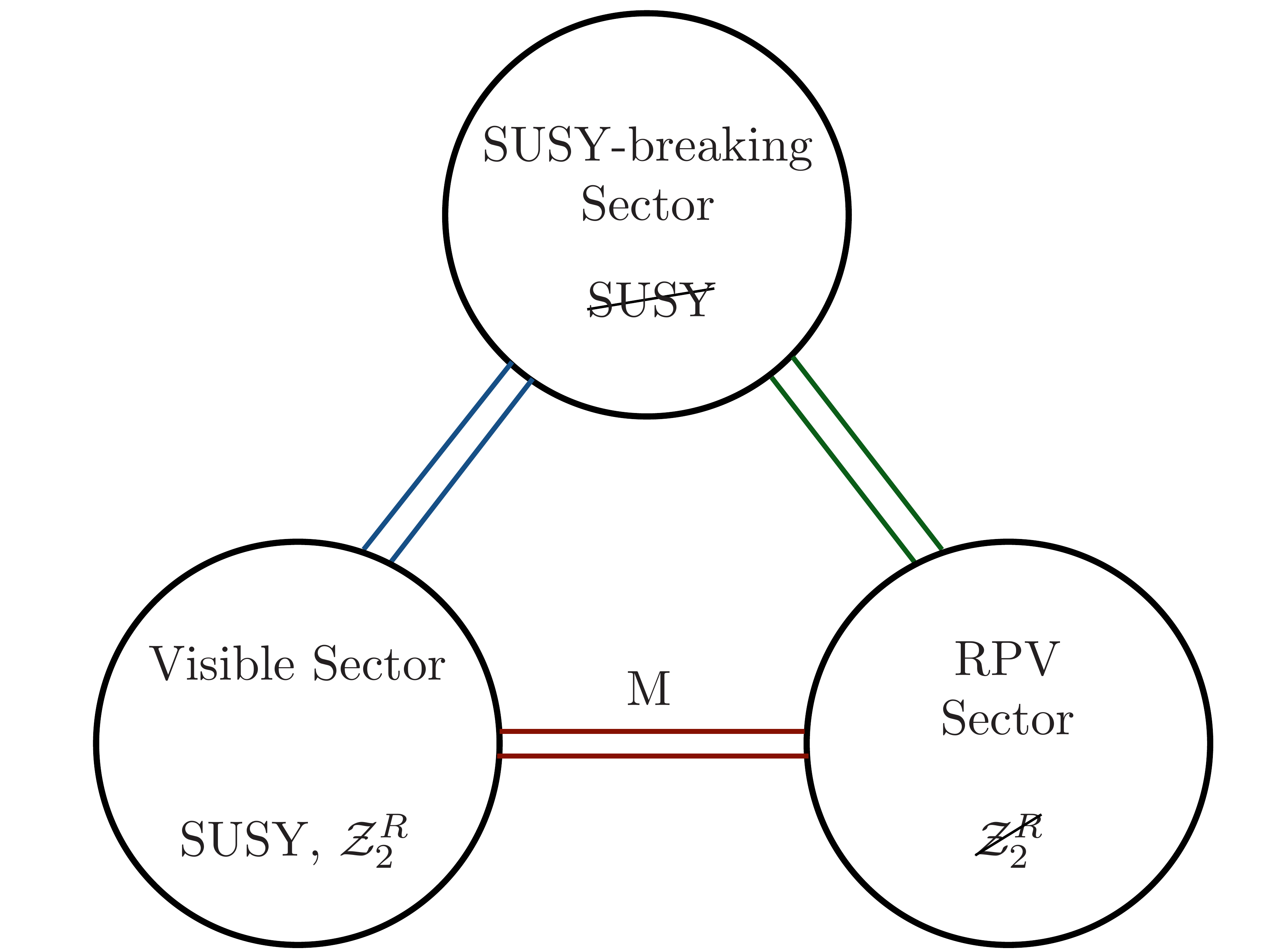}
\includegraphics[width=.46\textwidth]{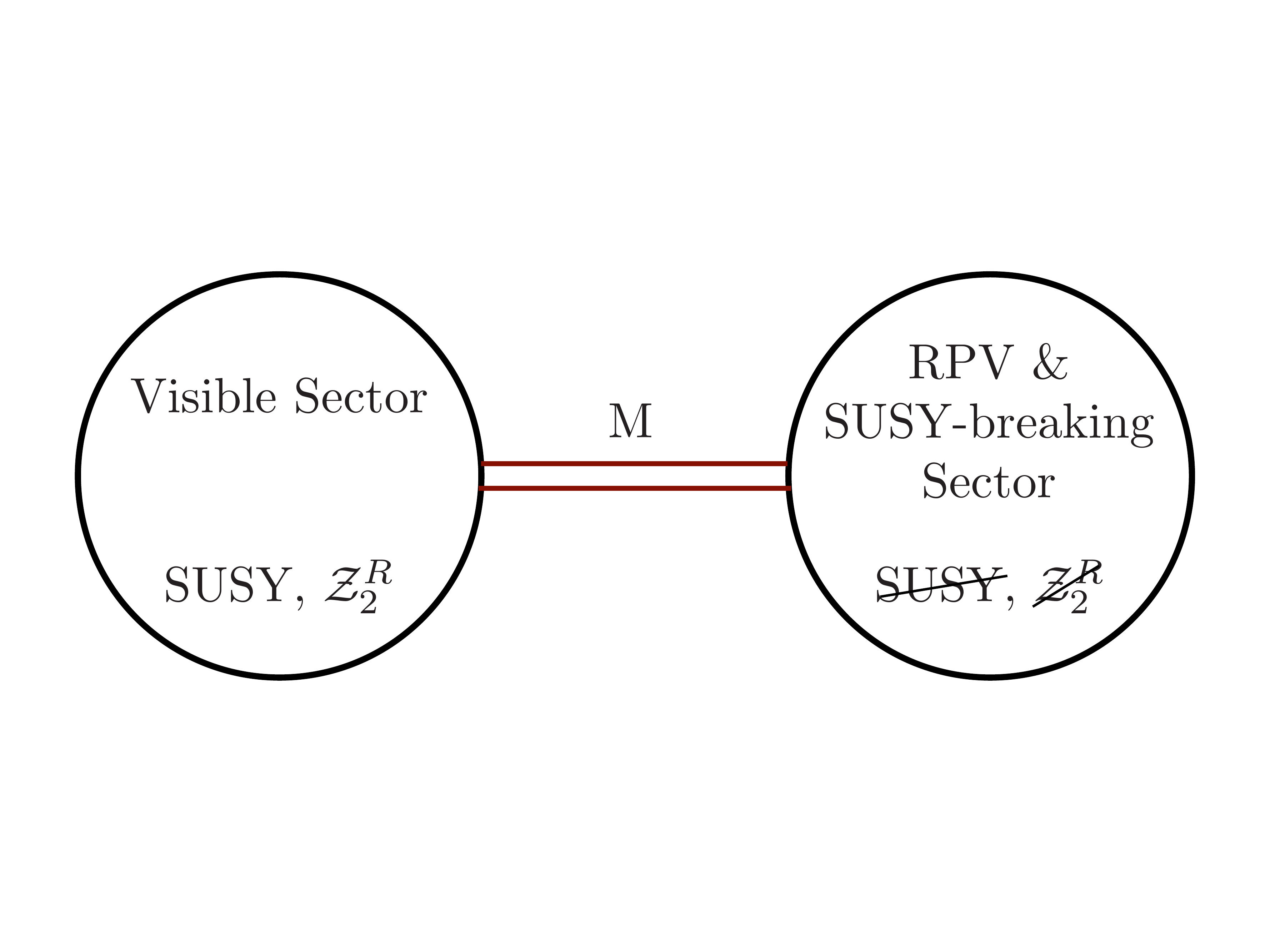}
\caption{\small{An illustration of the two SUSY-breaking scenarios considered.  {\bf Left:} The SUSY-breaking sector is secluded from both the visible sector and the R-parity breaking sector.   SUSY-breaking effects are then communicated to both sectors and show up as soft masses in the \kahler potential.   {\bf Right:} Both SUSY and R-parity breaking occur jointly in a single sector and are mediated to the visible sector at the scale $M$.}
\label{fig:sectors}}
\end{center}
\end{figure}

So far, our analysis has ignored any effects of supersymmetry breaking.  In this section, we analyze effects of SUSY-breaking on the low energy effective RPV terms. We  distinguish between two separate cases. (1) SUSY-breaking occurs in a secluded sector, different from that which breaks R-parity.  If the scale of SUSY-breaking is higher than the scale of RPV mediation, one may treat its effects by considering the expected soft breaking terms. (2)~SUSY-breaking and R-parity are broken in a single sector and both effects are mediated to the visible sector by the same mediator  fields.   These two scenarios are illustrated in Fig.~\ref{fig:sectors}.

\subsection{Secluded SUSY breaking Sector}

First we consider the scenario where SUSY breaking is external to the RPV sector.  Following the discussion in~\cite{Giudice:1997ni}, this situation can be described by incorporating the SUSY breaking spurion, $X= \langle X \rangle + \theta^2 F_X$, into  a wave-function renormalization factor of the RPV-messenger and visible sector fields,
\beq
\int d^4\theta  Z_i(X,X^*) \Phi_i \Phi^{i*} \label{susybreaking}\,.
\eeq
Here $\Phi_i$ is a generic field with $i$ running over the RPV-messenger and visible sector fields. 
Expanding  $  Z_i(X,X^*) $ in (\ref{susybreaking}) in superspace gives,
\beq
\int d^4\theta \left(  Z_i + \frac{\partial Z_i }{\partial X} F_X \theta^2 + \frac{\partial Z_i }{\partial X^*} F_X^* \bar{\theta}^2 + \frac{\partial^2 Z_i }{\partial X \partial X^*} |F_X|^2 \theta^2  \bar{\theta}^2  \right) \Phi_i \Phi^{i*} \,,
\eeq
where $ Z_i= Z_i(\langle X \rangle ,\langle X^* \rangle) $.  One can remove the  terms linear in $F_X$ and canonically normalize the  kinetic terms using the holomorphic field redefinition,
\beq
\Phi_i \rightarrow  Z_i^{-1/2} \left(1 - \frac{1}{  Z_i} \frac{\partial Z_i }{\partial X} F_X \theta^2 \right) \Phi_i \equiv z_i  \Phi_i \,.\label{susytransform}
\eeq
Under this transformation the original K\"ahler term (\ref{susybreaking}) becomes a canonical kinetic term along with standard soft breaking scalar masses,
\beq
 \int d^4\theta \left(1 + \frac{\partial^2 \log Z_i }{\partial X \partial X^*} |F_X|^2 \theta^2 \bar{\theta}^2 \right) \Phi_i \Phi^{i*}\,.
\label{diagK}
\eeq
At the same time, every operator in the theory must be  transformed according to Eq.~\eqref{susytransform}. 

Applying this transformation to the toy model of Sec.~\ref{sec:toymodel},  the effects of SUSY-breaking can be captured by  transforming $M$, $S$ and $\lambda$ as 
\beq
M \rightarrow M z_{D} z_{\bar{D}},~~~~~~~~ S \rightarrow S z_{D} z_{\bar{d}},~~~~~~~~ \lambda \rightarrow \lambda z_{q}^2 z_{D}\ .
\eeq
The effects of SUSY-breaking on the dRPV coupling (\ref{qqd}) are summarized in the operator
\beq
K_{dRPV} = \frac{z_{\bar{d}}^* }{z_{\bar{D}}^*}\frac{ z_{q}^2  }{ z_{\bar{D}}^{} } \frac{\lambda S^*}{|M|^2}  q q \bar{d}^* +h.c.\,. 
\label{drpvsusy}\,
\eeq
In addition to generating the chirally suppressed operators, SUSY-breaking RPV operators are also generated due to the F-components of the $z$-factors. In fact, when integrating over superspace, the $F^*$-term contributions from ${z_{\bar{d}}^* }/{z_{\bar{D}}^*} $ can dominate the  RPV effects over  the SUSY conserving ones. 

Let us estimate the magnitude of the SUSY-breaking dRPV operators. In many models of supersymmetry breaking and mediation, the wave-function renormalization factors, $Z_i$, can be approximated by the form,
\beq
Z(X,X^*) \sim 1- \frac{|X|^2}{\Lambda^2}\,.
\eeq
In gauge-mediated SUSY-breaking, $\langle X \rangle$ is the mass of the messengers and $\Lambda \sim 4 \pi \langle X \rangle / \alpha$, where $\alpha$ is the corresponding gauge coupling. In gravity mediation and anomaly mediation, $\Lambda \sim M_{\rm Pl}$ and $\Lambda \sim 4 \pi M_{\rm Pl}/\alpha$ respectively and, while $\langle X \rangle$ is in principle free, it is expected that $\langle X \rangle \ll M_{\rm Pl}$ in gravity mediation and $\langle X \rangle \simeq M_{\rm Pl}$ in anomaly mediation. From (\ref{diagK}), the soft-masses are
\beq
m_{\phi_i}^2 = -  \frac{\partial^2 \log Z_i }{\partial X \partial X^*} |F_X|^2 \sim m_0^2 \equiv \frac{|F_X|^2}{\Lambda^2} \,,
\eeq
while supersymmetry breaking suppressed dRPV operators will be proportional to 
\beq
\frac{F_{z_i^*}}{M} \sim \frac{\langle X^* \rangle F_X^*}{\Lambda^2 M} \sim \frac{\langle X^* \rangle}{\Lambda} \frac{m_0}{M}\,,
\eeq
which should be compared with the fermion mass, $m_f/M$, for chirally suppressed couplings. 
Therefore, in models of gravity  mediation the SUSY-breaking terms are typically negligible, but in  gauge and anomaly mediation, the SUSY-breaking terms could be much larger than the SUSY preserving (chirally suppressed) operators. 

If the scale of SUSY-breaking is below the scale of RPV mediation, $M$, then $z_{\bar{D}} \rightarrow 1$ and the suppression due to SUSY-breaking is $m_0 / 16\pi^2 M$. This suppression is expected to be less than the chiral suppression, with the exception of perhaps the operator containing the bottom quark. However, there can also be additional suppression to the SUSY-breaking couplings. For instance, in the case where the scale of gauge mediation is above the scale of the dRPV messengers, one finds at one-loop that $z_{\bar d} \sim z_{\bar D}$, since they have the same SM gauge quantum numbers. This implies that the effects of SUSY-breaking will be two-loop suppressed in the RPV couplings.

\subsection{Unifying dRPV and SUSY breaking}
\label{directsusy}

A second  possibility is that the messengers of dRPV couple directly to the SUSY-breaking sector.  In this case, they also become the messengers of SUSY-breaking. This scenario can easily be understood in the toy model of Sec.~\ref{sec:toymodel} by replacing $M\rightarrow X=M + \theta^2 F_X$, where $X$ is the SUSY-breaking spurion, so that the effective low energy K\"ahler potential becomes,
\begin{equation}
K_{eff}  =  |q|^2 + |\bar{d}|^2+ \frac{|S|^2}{|X|^2}  |\bar{d}|^2 +  \frac{\lambda S^*}{|X|^2}  q q \bar{d}^* + h.c. + \ldots   \label{toyeff2}
\end{equation}
The last term will contain SUSY-suppressed dRPV, which is larger than in the case where SUSY is broken in a different sector. However, the third term above can be  problematic, as it introduces a negative tree-level mass for the down-squarks (and the sleptons in a complete model). The magnitude of these negative scalar mass squares are of order
\beq
m_{\tilde{\ell},\tilde{\bar{d}}}^2 \simeq -\frac{\langle S \rangle^2}{M^2} \frac{F_X^2}{M^2}\ .
\label{eq:neg_soft_mass}
\eeq
Meanwhile, all the superpartners receive the usual positive mass squares from the corresponding mediation scheme.  For instance, in gauge mediation
\beq
m_{1/2}^2,m_0^2 \simeq \left(\frac{\alpha}{4\pi}\right)^2 \frac{F_X^2}{M^2}\ .
\eeq
All squared scalar masses  will be positive as long as $\langle S \rangle$ is not too large, namely $\langle S \rangle /M \lesssim \alpha_2 / 4\pi  \sim 10^{-3}$. In this scenario there is at least a factor of $10^{-3}$ overall suppression of all RPV couplings in addition to the chiral or SUSY-breaking suppression.  There can be additional flavor suppression which will be discussed in Sec.~\ref{sec:flavordrpv}. 

Negative soft masses may be avoided in models where R-parity is violated in a hidden sector.  Considering the hidden sector toy model of Sec.~\ref{sec:goodmodel}, but taking separate mass scales for the $D_+, \bar{D}_+$ and $D_-, \bar{D}_-$ fields,
\beq
W=M_+D_+ \bar{D}_+ +M_-D_-\bar{D}_- + S \bar{D}_+ D_- + m D_-\bar{d} + qq D_+\, , \label{improvedtoymplusminus}
\eeq
the effective low energy K\"ahler potential takes the form,
\beq
K_{eff} = |q|^2 + |\bar{d}|^2 + \frac{1}{|M_-|^2} \left(|m|^2 |\bar{d}|^2 + \frac{|S|^2}{|M_+|^2} |q|^4 - \frac{m^* S}{M_+} q q \bar{d}^* + h.c. \right) + ...
\eeq
In this case, only $D_-$ couples directly to the MSSM and to the RPV spurion.  Promoting $M_+ \rightarrow X$ to be the SUSY-breaking spurion induces a mediation of SUSY-breaking through  the $D_+, \bar{D}_+$ fields, but negative tree-level scalar masses are not generated.  Similarly, one may promote $S$ to the SUSY-breaking spurion, i.e. $S \rightarrow S + \theta^2 F_S$, obtaining a similar result.  
The  possibility in which the  same messenger fields mediate SUSY-breaking, flavor breaking and RPV, will be discussed in Sec.~\ref{sec:FlavorMediation}. 

\section{Flavor and Dynamical R-Parity Violation\label{sec:flavordrpv}}

So far  we have focused on the dynamics responsible for generating  the low energy non-holomorphic RPV operators, largely ignoring their flavor structure. However, flavor should not be ignored. The reason is that the R-parity violating interactions break the approximate $U(3)^5$ non-abelian global  flavor symmetries of the SM and may therefore induce flavor-violating interactions~\cite{Monteux:2013mna, Kong:1998bs}.   
For example, the model presented in~\eqref{toyfull} leads to dimension-6 four-fermi operators that may or may not respect the R-parity
and that violate baryon and lepton number.  
Consequently, such operators may, for instance, induce proton decay.  The chirally suppressed contribution predicts the proton lifetime to be,
\beq
\tau_p \simeq 10^{32} {\rm yr} \left( \frac{7 \times10^{-8}}{|\eta^{\prime\prime}_{ij3} \eta_{k\ell3}^*|} \right)^2 \left( \frac{{m}_{\tilde{b}_L}}{ \rm TeV} \right)^4 \left( \frac{M}{10^8 \rm GeV} \right)^4 \left( \frac{0.1}{\langle S \rangle /M } \right)^4.
\eeq
The full set of constraints is presented in Appendix~\ref{sec:FullModel}.
If the scale of dRPV mediation is sufficiently high, it may suppress such interactions without additional flavor suppression.  However, as a consequence, the LSP is predicted to be  collider-stable and the theory suffers from stringent collider constraints, pushing the superpartner's mass scale to be at around the TeV scale.  

One way out is  to introduce a non-trivial  flavor structure in the RPV operators. The main consequence of including flavor suppression is that they are not uniform, i.e. operators involving light fields will generically be more strongly suppressed than those involving the third generation, strongly improving bounds from baryon- and lepton-number violating processes, while at the same time allowing faster decays of the LSP (assuming it is a third generation superpartner). 
Interestingly, the SM flavor puzzle hints at the presence of a dynamical origin to the SM flavor parameters and any such dynamics  is expected to provide a correlated structure in the dRPV operators.  As presented below, known solutions to this puzzle easily provide the necessary flavor protection to dRPV operators.  
Moreover, since the fields that mediate RPV interact with the quark and leptons, assuming a flavor structure in the mediation scheme is a natural extension that may solve the SM flavor puzzle in conjunction with the dynamical generation of RPV operators.

In accordance with the above, we consider two possible approaches. First, we simply extend the dRPV model to include flavor suppression factors, assuming that the flavor physics occurs externally to the sector responsible for RPV (or at a higher scale). We then consider the case where the dRPV mediation sector itself is responsible for the generation of the SM flavor hierarchies. 
In principle, any solution to the SM flavor puzzle may be considered for these approaches.   For concreteness,  we consider the specific mechanism suggested by Froggatt and Nielsen~\cite{Froggatt:1978nt}, demonstrating the general principles.   Other solutions may be straightforwardly applied.

\subsection{An External Flavor Sector}

There are numerous models that attempt to solve the Standard Model flavor puzzle~\cite{De_Rujula:1977ry,Dimopoulos:1991yz,Cahn:1980kv,Balakrishna:1988bn,Kaplan:1991dc,KerenZur:2012fr, Agashe:2004cp, Agashe:2004rs, Nomura:2007ap}. Some of these solutions induce hierarchical structure within the RPV operators.   One well studied and motivated setup is known as the Froggatt-Nielsen (FN) mechanism which assumes that the hierarchy in the fermion masses and mixing angles are the result of the spontaneous breaking of horizontal (flavor dependent) $U(1)_{\rm FN}$ symmetries.  The symmetry breaking is communicated to the low energy theory at a scale $M_{\rm FN}$ in a flavor-dependent manner, dictated by the charges of the various SM fields.  The corresponding low energy operators are suppressed by different powers of the small parameter,
\beq
\epsilon \equiv \frac{\langle \phi \rangle}{M_{\rm FN}}\,,
\eeq
where $\phi$ is the Froggatt-Nielsen field(s) responsible for the breaking of the horizontal symmetry.
In particular,  the low energy RPV couplings of~\eqref{Kdrpv2} will be suppressed in a flavor dependent manner, 
\beq
\alpha, \eta,\lambda \propto \epsilon^{\mathcal{Q}} \,,
\eeq
where $\mathcal{Q}$ is the $U(1)_{\rm FN}$ charge of the operator. The FN charges of the SM fields are determined such that the correct fermion masses and mixing angles are reproduced, typically implying that the lighter generations have a larger FN charge. Therefore, the  RPV couplings of operators containing first generation fields will be more suppressed than the couplings of the heavier generations. The actual suppression factors may also depend on the FN charges of the heavy fields $D,\bar{D},L,\bar{L}$. 

Assuming that the heavy fields are uncharged under the FN symmetry, the flavor suppression of the dRPV operators of the model in Eq.~(\ref{Kdrpv2}) are,
 \begin{eqnarray}
\eta_{ijk} \propto \epsilon^{|\mathcal{Q}_{u_i}+\mathcal{Q}_{\bar{e}_j}|+|\mathcal{Q}_{\bar{d}_k}|}\ , \ \ 
&&\eta'_{ijk} \propto \epsilon^{|\mathcal{Q}_{q_i}+\mathcal{Q}_{\bar{u}_j}|+|\mathcal{Q}_{\ell_k}|}\ , \ \ \eta''_{ijk} \propto \epsilon^{|\mathcal{Q}_{q_i}+\mathcal{Q}_{q_j}|+|\mathcal{Q}_{\bar{d}_k}|} \,, \nonumber \\
&&\lambda^{\not{B},\not{L}}_{ijmn} \propto \epsilon^{|\mathcal{Q}_{q_i}+\mathcal{Q}_{q_j}|+|\mathcal{Q}_{\bar{u}_m}+\mathcal{Q}_{\bar{e}_n}|}\,.
\label{eq:FNsuppressions}
\end{eqnarray}
A particularly successful example of FN charges from \cite{Leurer:1992wg, Leurer:1993gy} is presented in Table~\ref{tab:FNcharges}, along with the coupling suppression factors in Tables~\ref{tab:etaetapFN} and \ref{tab:lambdaFN} of 
Appendix \ref{App:FNsuppression}.

\subsection{Unifying dRPV and Flavor Models}
\label{sec:FlavordRPV}

Next, we consider the possibility that both the RPV and the breaking of the flavor symmetry occur in a single sector and are mediated by the same set of messengers.  In this scenario, the RPV spurion, $S$, may also identified with the Froggat-Nielsen spurion. To illustrate this, consider adding the operator $h_d q \bar{D}$ to  the toy model superpotential (\ref{toymodel})
\beq
W = X  D  \bar{D}+ S  D  \bar{d} +\lambda  q  q D + h_d q \bar{D}\ ,
\label{toyfn}
\eeq
where $M$ has now been promoted to an R-parity odd spurion $X$ with $\left<X\right> = M$, since in this case the $\bar{D}$ field is now R-parity odd as opposed to the original toy model (\ref{toymodel}).  There is now a horizontal symmetry, with $\mathcal{Q}_S = -1$ and $\mathcal{Q}_{\bar{d}} =+1$, that forbids the standard Yukawa coupling $h_d q \bar{d}$. However, when integrating out the heavy fields, the low energy effective superpotential now contains the term,
\beq
 \frac{S}{X}h_d q \bar{d},
 \eeq
and the Yukawa coupling is generated with the suppression,
\begin{equation}
\epsilon \equiv \frac{\langle S\rangle}{X}\,.
\end{equation} 
In addition, the dRPV K\"ahler terms as in (\ref{qqd}) are generated, with the appropriate power of flavor suppression. 

In a similar fashion,the hidden sector model of (\ref{improvedtoy}) can be extended to mediate flavor, by considering the Lagrangian,
\begin{equation}
W=M(D_+ \bar{D}_+ +D_-\bar{D}_-) + S \bar{D}_+ D_- + m D_-\bar{d} + qq D_++h_d q \bar{D}_-,\,  \label{hiddenFN}
\end{equation}
and promoting the $D_-,\bar{D}_-$ to the FN messenger fields. The main advantage of this scenario is that no additional source for R-parity violation needs be introduced.  We return to this model in Sec.~\ref{sec:FlavorMediation} when we discuss the flavor mediation scenario. 

\begin{figure}[t!]
\begin{center}
\begin{fmffile}{matter}
    \begin{fmfgraph*}(130,30)
        \fmftop{t1,t2,t3,t4}
	\fmfleft{i1,i2}
       	\fmfright{o1,o2}

	\fmf{fermion}{i1,v1} \fmflabel{$q^{(k)}$}{i1}
	\fmf{fermion}{i2,v1}	\fmflabel{$h_d$}{i2}
	\fmf{fermion}{o1,v6} \fmflabel{$\bar{d}^{(j)}$}{o1}
	\fmf{dashes_arrow}{o2,v6} \fmflabel{$S^{(-1)}$}{o2}
	\fmf{fermion,label=$\bar{D}^{(-k)}$,l.side=left}{v2,v1}	
	\fmf{fermion,label=${D}^{(k)}$}{v2,v3}		
	\fmf{dashes_arrow}{t2,v3} 
	\fmf{fermion,label=$\bar{D}^{(-k+1)}$,l.side=left}{v4,v3}	
	\fmf{dots}{v4,v5}
	\fmf{fermion,label=${D}^{(-j+1)}$}{v5,v6}
	\fmfforce{.25w,.5h}{v1}
	\fmfforce{.35w,.5h}{v2} \fmfiv{d.sh=cross,d.ang=90,d.siz=5thick}{(.35w,.5h)}  
   \fmfforce{.38w,.6h}{l1} \fmflabel{$M$}{l1} 
	\fmfforce{.45w,.5h}{v3}
	\fmfforce{.55w,.5h}{v4} \fmfiv{d.sh=cross,d.ang=90,d.siz=5thick}{(.55w,.5h)} 
   \fmfforce{.525w,.55h}{l2} \fmflabel{$M$}{l2} 
	\fmfforce{.65w,.5h}{v5} \fmfiv{d.sh=cross,d.ang=90,d.siz=5thick}{(.65w,.5h)}
   \fmfforce{.625w,.60h}{l3} \fmflabel{$M$}{l3} 
	\fmfforce{.75w,.5h}{v6}
	\fmfforce{.35w,.9h}{t1}
	\fmfforce{.45w,.88h}{t2}  \fmfforce{.475w,.9h}{l4} \fmflabel{$S^{(-1)}$}{l4}
	\fmfforce{.55w,.9h}{t3}
	\fmfforce{.65w,.9h}{t4}	
	\fmfforce{.1w,.1h}{i1} 
	\fmfforce{.1w,.9h}{i2}
	\fmfforce{.9w,.1h}{o1}
	\fmfforce{.9w,.9h}{o2}	

   \end{fmfgraph*}
\end{fmffile}
\caption{
\label{fig:FN}
Generation of down-type Yukawa couplings in FN models via integrating out heavy vector-like fields. Superscripts denote the U(1)$_{\rm FN}$ charge.}
\end{center}
\end{figure}
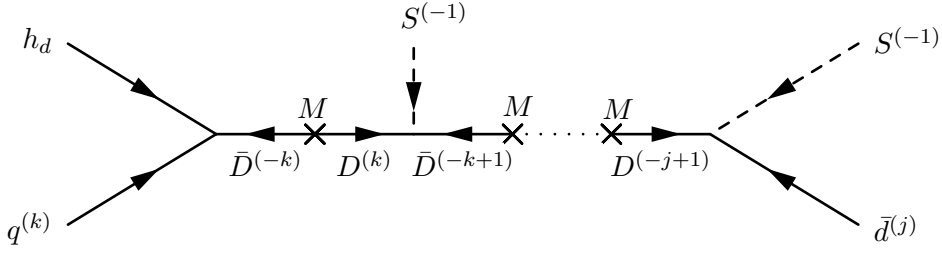

The above models can easily be extended to generate the full down-quark Yukawa matrix, with elements suppressed as  
\beq
Y^d_{ij} =\epsilon^{\mathcal{Q}_{\bar{d}_i}+ \mathcal{Q}_{q_j}}  \times \mathcal{O}(1),
\eeq
by including the full generation of quarks, and multiple $D, \bar{D}$ (or $D_-, \bar{D}_-$) messengers that carry different FN charges.  A generic diagram corresponding to the generation of the down-type Yukawa couplings is shown in Fig.~\ref{fig:FN}.   In the diagram, ${D}^{(i)}$ and $\bar{D}^{(i)}$ correspond to vector-like messengers with FN charge $i$.  
Upon integrating out the heavy fields, in addition to the appropriate suppression  of the Yukawa matrices, the dRPV terms will also be suppressed by powers of $\epsilon$. An advantage of this unified scheme is that, unlike the case where the flavor dynamics is external, the scale of the RPV spurion is set by flavor physics $\langle S\rangle/X\sim {\mathcal O}(0.2)$.

In order to generate masses for the up-type quarks, additional heavy up-quark singlets ($U,\bar{U}$) or doublets ($Q,\bar{Q}$) must be included. For simplicity, whole heavy vector-like generations ($Q_i,\bar{U}_i,\bar{D}_i,L_i,\bar{E}_i$) can be introduced.  This could be motivated by the preservation of gauge coupling unification. Generalizing the previous model (\ref{toyfull}) to include these new fields one finds,
  \beq\begin{array}{rcl}
W &=& X \left( Q \bar{Q}+  D \bar{D}+  U \bar{U}+ L \bar{L} +E \bar{E}  \right)+S \left(Q \bar{Q}+  D \bar{D}+   U \bar{U}+ L \bar{L} +   E \bar{E} \right)\\
&&+S \left(q \bar{Q}+  D \bar{d}+   U \bar{u}+ \ell \bar{L} +   E \bar{e} \right)+h_u Q \bar{U}+h_d Q \bar{D}+h_d L \bar{E} \\
&&+ \bar{u} \bar{e} D + Q u \bar{L} +  q q D\\
&&+ \mu h_u h_d \,,
 \label{drpvfnfull}
\end{array}\eeq 
where we have suppressed all (${\cal O}(1)$) flavor indices and coupling constants.\footnote{Note that the above superpotential is most easily discussed in SU(5) language:
  \[\begin{array}{rcl}
W &=&  X ({\bf 10_H \overline{10}_H +  5_H \bar{5}_H }) +  S ({\bf 10_H \overline{10}_H +  5_H \bar{5}_H } )+S({\bf  \overline{10}_H 10_L +  5_H \bar{5}_L } )\\
&&+\,  h_u {\bf \overline{10}_{H,L}  \overline{10}_{H,L} } + h_d {\bf \overline{10}_{H,L} \bar{5}_{H,L} } 
+ {\bf 10_L 10_L {5}_L}
+ \mu h_u h_d \,.
\label{drpvfnfullsu5}
\end{array}\] }
The form of  suppression of the dRPV operators is slightly different from the case where the FN mechanism is external to the RPV sector, since the heavy fields are now charged under the FN symmetry.  The flavor suppression is now,
\begin{eqnarray}
\eta_{ijk} \propto \epsilon^{|\mathcal{Q}_{u_i}+\mathcal{Q}_{\bar{e}_j}-\mathcal{Q}_{\bar{d}_k}|}\ , \ \ && 
\eta'_{ijk} \propto \epsilon^{|\mathcal{Q}_{q_i}+\mathcal{Q}_{\bar{u}_j}-\mathcal{Q}_{\ell_k}|}\ , \ \ \eta''_{ijk} \propto \epsilon^{|\mathcal{Q}_{q_i}+\mathcal{Q}_{q_j}-\mathcal{Q}_{\bar{d}_k}|}\,, \nonumber \\
&&\lambda^{\not{B},\not{L}}_{ijmn} \propto \epsilon^{|\mathcal{Q}_{q_i}+\mathcal{Q}_{q_j}-\mathcal{Q}_{\bar{u}_m}\mathcal{Q}_{\bar{e}_n}|}\,.
\label{eq:FNsuppressionsinternal}
\end{eqnarray}
The same set of FN charges given in Table~\ref{tab:FNcharges} is also consistent with this scenario.

\section{Towards dRPV Flavor Mediation} 
\label{sec:FlavorMediation}

We close the paper by discussing a scenario in which the  breaking of supersymmetry, R-parity and FN horizontal symmetry is mediated jointly  by the same sets of heavy messengers.   This framework is a deformation of the well-studied flavor mediation scenario in which supersymmetry breaking is mediated in a flavor-dependent manner~\cite{Kaplan:1998jk, Kaplan:1999iq, Shadmi:2011hs, Abdullah:2012tq, Galon:2013jba, Nomura:2008gg, Nomura:2008pt, Craig:2012di, Pomarol:1995xc, Grossman:2007bd}.  Below we describe the main features and problems of such a scenario, postponing a detailed study to a future publication.  

Let us begin by considering the toy model~\eqref{toyfn} or its extension~\eqref{drpvfnfull}, described in Sec.~\ref{sec:FlavordRPV}.  Following  the discussion in Sec.~\ref{directsusy}, we promote the field, $X$, to be the SUSY-breaking spurion, $X\equiv M + \theta^2 F$.  In order to account for the fermion masses, $\epsilon=\langle S \rangle / M$ is typically chosen to be of order ${\cal O}(0.2)$ although smaller values may be considered.   
As shown in Sec.~\ref{directsusy}, two contributions to the soft masses in the visible sector are obtained once the mediators are integrated out.  The first are the usual (positive) gauge-mediated contributions.  The second type arising at tree-level from the direct flavor-dependent couplings, are negative and of order $\tilde m^2 \simeq  (\langle S \rangle /M)^2 F_X^2/M^2$, see Eq.~\eqref{eq:neg_soft_mass}.  Unless $\epsilon$ is sufficiently small, the resulting spectrum is not viable. 

There are several  possible avenues to eliminate the tension described above.  One possibility is to promote one of the global U(1) symmetries, such as $B-L$, to a local U(1) gauge symmetry.  In this scenario, all scalars receive an additional contribution to their mass of order
\beq
m_0^2 \simeq \left(\frac{\alpha_{B-L}}{4\pi}\right)^2 \frac{F_X^2}{M^2}.
\eeq
For $\alpha_{B-L} \gtrsim 1$, the above may dominate over the negative contributions of Eq.~\eqref{eq:neg_soft_mass}. However, since the gauginos still acquire mass only from the SM gauge mediation, this scenario predicts a hierarchy between the scalars and the gauginos,
\beq
m_0 \simeq \frac{\alpha_{B-L}}{\alpha_i} m_{\chi_i}  \,,
\eeq
 for $i = 1,2,3$.  In particular, one finds that the stop mass is significantly larger than the bino mass, with $m_{\tilde{t}} \gtrsim 10 \cdot  m_{\tilde{B}}$.  Since $m_{\tilde{B}}\geq 100$ GeV, we find $m_{\tilde{t}}\geq 1$ TeV, which introduces sizable (of order percent) tuning into the model. 
 
Another possibility is to introduce additional messengers between the visible and dRPV sector at a scale $M_*$, which is below the dRPV messenger scale $M$. This will induce positive masses of order
\beq
m_0^2 \simeq \left(\frac{\alpha}{4\pi}\right)^2 \frac{F_X^2}{M_*^2},
\eeq
so provided that $M_*/M \lesssim \frac{\alpha_2 / 4\pi}{\epsilon} \simeq 10^{-2}$, the scalar masses will remain positive.

The above discussion is relevant for an extension of the original  toy model (\ref{toymodel}) presented in Sec.~\ref{sec:toymodel}.   A more appealing framework for dRPV flavor mediation is that based on the hidden sector model (\ref{improvedtoy}) presented in Sec.~\ref{sec:goodmodel}.  Indeed, as discussed in Sec.~\ref{directsusy}, the simple extension of the hidden sector model given in Eq.~\eqref{improvedtoymplusminus} allows for the introduction of supersymmetry breaking without inducing any negative masses.    The FN mediation and the corresponding generation of the SM Yukawa couplings follows by promoting $m\rightarrow \phi$ to be the FN breaking spurion (as opposed to the previous case in which $S$ was responsible for the horizontal breaking) and adding the corresponding terms which generate the Yukawa interactions.  Explicitly, the hidden sector toy model takes the form,
\beq
W=M_+ D_+ \bar{D}_+ +M_-D_-\bar{D}_- + S \bar{D}_+ D_- + \phi D_-\bar{d} + qq D_+ + h_d q \bar D_- \, ,
\eeq
where $S = \langle S\rangle + \theta^2 F_S$ is responsible for breaking both SUSY and R-parity. 
Following the discussion in Sec.~\ref{sec:FlavordRPV}, extending this model to a complete one, including all SM baryons and leptons, is straightforward and follows from the introduction of family of $D_{-}, \bar{D}_{-}$ mediators which carry different FN charges.  The $D_+, \bar{D}_+$ mediators do not play a direct role in the Yukawa structure of the effective low energy theory.  Finally, the spectrum of the visible sector superpartners is the usual gauge mediated spectrum.   
A more thorough study of this model and its phenomenology is beyond the scope of this work.

\section{Conclusions\label{sec:Conclusions}}
In this work we have explored theories of dynamical R-parity violation, in which the breaking of R-parity is communicated to the visible sector by heavy mediator fields.  These models address the origin of R-parity violation and naturally explain the smallness of the RPV effects.   As was originally shown  in~\cite{Csaki:2013jza}, dynamical RPV predicts novel, non-holomorphic RPV operators, Eq.~\eqref{eq:KRPV}, 
that  often dominate over the standard holomorphic interactions typically considered in the literature. The operators exhibit phenomenology distinct from the standard scenario and may have unique signatures at experiments such as the LHC.  A future publication~\cite{Third_Paper:2015} will present the phenomenology expected from these models and suggest explicit search strategies  for the $\sqrt{s}=14\text{ TeV}$ run. 

The models presented here exhibit an R-parity symmetry which is spontaneously broken in the UV and is communicated to the visible sector via a set of heavy messenger fields. The effects of the non-holomorphic dRPV operators are  suppressed by the ratio of either the light fermion masses or the SUSY breaking scale to the scale of RPV mediation.  In addition we have argued that any theory which addresses the hierarchical flavor structure of the SM will directly affect the structure of the RPV operators, leading to additional flavor dependent suppressions.  
The overall result is that the dRPV operators can be sufficiently suppressed to evade all experimental bounds including proton decay and neutrino masses, and that the low energy theory exhibits novel LHC phenomenology.

We have studied the possibility of  unifying the generation of both the SM flavor structure and the R-parity violating interactions within the framework of the Froggatt-Nielsen mechanism.   
Other mechanisms, such as partial compositeness, are expected to be equally compatible with the dRPV framework.  
We have also considered the possibility that the breaking of supersymmetry is mediated in conjunction with the mediation of RPV.  
Finally, we have shown how to extend these simple models such that
the heavy fields which mediate RPV also mediate the SUSY-breaking effects and the flavor structure, resulting in a flavor-mediation model which features the dynamical generation of dRPV, SUSY-breaking and flavor structure of the low energy effective theory.  

\section*{Acknowledgments}
We thank Sal Lombardo, Stefan Pokorski and Robert Ziegler for useful discussions.  C.C. thanks the members of the Particle Physics Group at Tel Aviv University for their hospitality while this work was in progress. C.C. and E.K. are supported in part by the NSF grant PHY-1316222. O.S. and T.V. are supported in part by a grant from the Israel Science Foundation.  T.V. is further supported by  the US-Israel Binational Science Foundation, the EU-FP7 Marie Curie, CIG fellowship and by the I-CORE Program of the Planning Budgeting Committee and the Israel Science Foundation (grant NO 1937/12).

\appendix

\section{Phenomenological Aspects of a Complete Model}
\label{sec:FullModel}
This section is devoted to summarizing the consequences of the main constraints on these models arising from proton decay, neutrino masses and the LSP lifetime. For simplicity we will use the minimal realization of dRPV based on the superpotential (\ref{toyfull}), introduced in Section \ref{sec:complete}. 

\subsection{Proton Decay}
Combinations of baryon and lepton violating operators will contribute to proton decay~\cite{Nakamura:2010zzi, Nishino:2009aa, Kobayashi:2005pe}.  The K\"ahler terms in Eq. (\ref{Kdrpv2}) which are relevant, i.e. violate either $B$- or $L$-number or both, are,
\beq
K_{\not{B},\not{L}} =\frac{{\lambda_{ijmn}^{\not{B},\not{L}}}}{2 M^2}q_iq_j \bar{u}^*_m\bar{e}^*_n + \frac{\langle S^* \rangle}{M^2}
  \left( \eta_{ijk}\bar{u}_i\bar{e}_j\bar{d}^{*}_k + \eta^\prime_{ijk}  q_i\bar{u}_j \ell^{*}_k + \frac{1}{2}{\eta^{ \prime\prime}_{ijk}}  q_i q_j \bar{d}^{*}_k \right)+ h.c.
\label{eq:K_Matter_Model}
\eeq
One can take this to be the starting point for the phenomenology and use the coefficients as free parameters. 
The simplest way to capture the effects of supersymmetry breaking in the low energy Lagrangian is to promote the $\eta$'s to superfields  
\beq
{\eta_{ijk}} \rightarrow {\eta_{ijk}}   +  {\eta_{ijk}^{\bar{\theta}^2}}  \bar{\theta}^2 \ .
\eeq
To estimate the lifetime of the proton, one can integrate out the superpartners and find the coefficients of the effective four-fermion operators that violate both $B$- and $L$-number,
\beq \arraycolsep=1.4pt\def\arraystretch{2.0}\begin{array}{rcl}
\mathcal{L}_{\not{B},\not{L}} 
=&&\Big[  \frac{|\langle S \rangle|^2}{2M^4} \frac{1}{m_{\tilde{d}_{L,k}}^2} \left(m^d_j (\eta^{ \prime\prime}_{ikj} + \eta^{ \prime\prime}_{kij}) + m^d_k (\eta^{ \prime\prime}_{ijk} + \eta^{ \prime\prime}_{jik}) \right) \left( m^d_k \eta^{*}_{mnk} + m^e_n \eta^{\prime *}_{kmn}\right)\\
&&~+  \frac{|\langle S \rangle|^2}{2M^4}\frac{1}{m_{\tilde{d}_{R,i}}^2}  ( \eta^{ \prime\prime\,{\bar{\theta}^2} }_{ijk} + \eta^{ \prime\prime\,{\bar{\theta}^2} }_{jik}) \eta^{{\bar{\theta}^2} *}_{mnk} + \frac{1}{2M^2 } ( {\lambda_{ijmn}^{\not{B},\not{L}}} + {\lambda_{jimn}^{\not{B},\not{L}}}) \Big]  u_{L}^i d_{L}^j u_{R}^{m} e_{R}^n +h.c. \\
\equiv&&{\tilde{\Lambda}^{-2}_{ijmn}} u_{L}^i d_{L}^j u_{R}^{m} e_{R}^n + h.c. 
\end{array}
\label{eq:pdecay4fermi}\eeq
with
\beq\arraycolsep=1.4pt\def\arraystretch{2.0}\begin{array}{c}
\eta^{ \prime\prime\,{\bar{\theta}^2} }_{ijk}  =  \eta^{\prime\prime}_{ijk} \left(\frac{\partial Z_{\bar{D}}}{\partial X^*} - \frac{\partial Z_{\bar{d}_k}}{\partial X^*} \right)F_X^* \,, ~~~~~~~
\eta^{{\bar{\theta}^2}}_{ijk}  =  \eta^*_{ijk} \left(\frac{\partial Z_{\bar{D}}}{\partial X^*} - \frac{\partial Z_{\bar{d}_k}}{\partial X^*} \right)F_X^* \,,\\
\eta^{ \prime\,{\bar{\theta}^2} }_{ijk}  =  \eta^{\prime}_{ijk} \left(\frac{\partial Z_{\bar{L}}}{\partial X^*} - \frac{\partial Z_{\bar{\ell}_k}}{\partial X^*} \right)F_X^*.
\end{array}\eeq
The first set of terms contains the chirally suppressed operators, the second set arises from SUSY-breaking, while the third set contains the four-Fermi operators that conserve R-parity. From these operators, the proton can decay via $p \rightarrow (\pi \rm{~or~} K) (e^+ \rm{~or~} \mu^+) $. 

The matrix element for the process is 
\beq
\mathcal{M}_{ijmn} \simeq  \left(\frac{\tilde{\Lambda}_{QCD}}{\tilde{\Lambda}_{ijmn}}\right)^2\ ,
\eeq
where $\tilde{\Lambda}_{QCD}$ is a nuclear matrix element expected to be of the order of $\Lambda_{QCD} \sim 200$ MeV~\cite{Aoki:2008ku}. The resulting proton decay width is
\beq
\Gamma_{ijmn} \simeq  \frac{m_p}{8\pi} \left| \mathcal{M}_{ijmn}  \right|^2,
\eeq
corresponding to a  lifetime of
\beq
\tau \simeq 10^{32} {\rm yr} \left(\frac{250 {\rm ~MeV}}{\tilde{\Lambda}_{QCD}} \right)^{4}  \left(\frac{\tilde{\Lambda}_{ijmn} }{10^{15} {\rm GeV}} \right)^4.
\eeq
\begin{figure}[t!]
\begin{center}
\begin{fmffile}{pdecay1a}\hspace*{2mm}
	        \begin{fmfgraph*}(15,15)
	     \fmfstraight
	        \fmfleft{i1,i2}
	        \fmfright{o1,o2}
                 \fmf{fermion}{i2,v1}
                 \fmf{fermion}{i1,v1}
                 \fmfblob{.10w}{v1}
                 \fmf{fermion}{v1,o2}
                 \fmf{fermion}{v1,o1}

                  \fmflabel{$d_L$}{i1}
	        \fmflabel{$u_L$}{i2}
	        \fmflabel{$u_R$}{o1}
	        \fmflabel{$e_R$}{o2}
 
	        \end{fmfgraph*}
\end{fmffile}
\hspace{2cm}
\begin{fmffile}{pdecay1b}\hspace*{2mm}
	        \begin{fmfgraph*}(30,15)
	     \fmfstraight
	        \fmfleft{i1,i2}
	        \fmfright{o1,o2}
                 \fmf{fermion}{i2,v1}
                 \fmf{fermion}{i1,v1}
                 \fmf{dashes,label=$\tilde{d}$}{v1,v2}
                 \fmf{fermion}{v2,o2}
                 \fmf{fermion}{v2,o1}
                  \fmflabel{$d_L$}{i1}
	        \fmflabel{$u_L$}{i2}
	        \fmflabel{$u_R$}{o1}
	        \fmflabel{$e_R$}{o2}
	        \end{fmfgraph*}
\end{fmffile}
\end{center}
\caption{Diagrams contributing to proton decay.}
\label{Fig:neutrinomasses}
\end{figure}
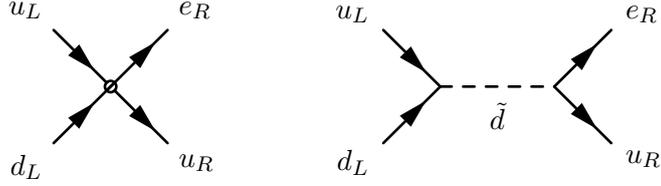
We can then present estimates for the proton lifetime assuming one of the three contributions in (\ref{eq:pdecay4fermi}) dominates. Of course in the generic case all three terms contribute and interfere with each other. 

First we assume that the chirally suppressed operators dominate. The leading contribution from these operators is expected to arise from the exchange of a sbottom squark, because in this case the chiral suppression can be given by the bottom mass (instead of the down or strange masses). Assuming this term indeed dominates, the proton lifetime is estimated to be,
\beq
\tau_p \simeq 10^{32} {\rm yr} \left( \frac{7 \times10^{-8}}{|\eta^{\prime\prime}_{ij3} \eta_{mn3}^*|} \right)^2 \left( \frac{{m}_{\tilde{b}_L}}{ \rm TeV} \right)^4 \left( \frac{M}{10^8 \rm GeV} \right)^4 \left( \frac{0.1}{\langle S \rangle /M } \right)^4 .
\eeq

Next we consider the case where the leading contribution comes from the SUSY-breaking terms. To simplify the discussion, we assume that the supersymmetry breaking effects are universal and flavor blind. In this case, we can parametrize the effects of SUSY-breaking by introducing the parameter $\epsilon_X$,
 \beq
\eta \equiv\eta^0(1+M\epsilon_X\bar{\theta}^2). \label{susyepsilonx}
\eeq
$\epsilon_X$ is of order $F_{z_i}/M \sim \frac{\langle X\rangle m_0}{\Lambda M}$ for the case when SUSY-breaking is external to the dRPV sector and $\epsilon_X \sim \frac{F_X}{M^2}$ for the case when SUSY-breaking is directly coupled. The proton lifetime for this case is given by
\beq
\tau_p \simeq 10^{32} {\rm yr} \left( \frac{10^{-8}}{|\eta^{\prime\prime}_{ijk} \eta_{mnk}^*|} \right)^2 \left( \frac{{m}_{\tilde{d}_{L,k}}}{ \rm TeV} \right)^4 \left( \frac{10^{-7}}{\epsilon_X} \right)^4 \left( \frac{0.1}{\langle S \rangle /M } \right)^4 .
\eeq
For a model of minimal gauge mediation occurring below the scale of the RPV mediation, one finds,
\beq
\left( \frac{{m}_{\tilde{d}_{L,k}}}{ \rm TeV} \right) \left( \frac{10^{-7}}{\epsilon_X} \right) \simeq \left( \frac{M}{10^{8} \rm GeV} \right).
\eeq

Finally the proton can decay, unsuppressed by SUSY-breaking, RPV, or fermion masses via the final baryon- and lepton-number violating but R-parity conserving 4-Fermi operator. If this dominates, the lifetime is approximately given by
\beq
\tau_p \simeq 10^{32} {\rm yr} \left( \frac{10^{-14}}{|\lambda^{\not{B},\not{L}}_{ijmn}|} \right)^2  \left( \frac{M}{10^8 \rm GeV} \right)^4.
\eeq

\subsection{Neutrino Masses}
\begin{figure}[h!]

\begin{center}
\begin{fmffile}{neutrino1}\vspace*{15mm}
  \begin{fmfgraph*}(40,12)
 	        \fmfsurroundn{i}{4}
                 \fmfforce{(0.5w,1.7h)}{i2}
                 \fmfforce{(0.5w,-0.7h)}{i4}
                 \fmflabel{$\ell$}{i1}
                 \fmflabel{$\ell$}{i3}
                 \fmflabel{$h_u$}{i2}
                 \fmflabel{$h_u$}{i4} 
                 \fmf{fermion}{i3,v1}
                 \fmf{fermion}{i1,v2}
                 \fmf{phantom,left,tension=0.25,tag=1}{v1,v2}
                 \fmf{phantom,left,tension=0.25,tag=2}{v2,v1}
                 \fmfposition
                 \fmfipath{p[]}
                 \fmfiset{p1}{vpath1(__v1,__v2)}
                 \fmfiset{p2}{vpath2(__v2,__v1)}
                 \fmfi{dashes_arrow,label=$\tilde{\bar{u}}$}{subpath (0,length(p1))/2 of p1}
                \fmfi{dashes_arrow,label=$\tilde{q}$}{subpath (length(p1),length(p1)/2) of p1}
                  \fmfi{fermion,label=$\bar{u}$}{subpath (0,length(p2)/2) of p2}
                 \fmfi{fermion,label=$q$}{subpath (length(p2),length(p2)/2) of p2}
                 \fmfi{dashes_arrow}{(0.5w,1.7h) -- point length(p1)/2 of p1}
                \fmfi{dashes_arrow}{(0.5w,-0.7h) -- point length(p2)/2 of p2}
 
  \end{fmfgraph*}
  \hspace*{25mm}
	        \begin{fmfgraph*}(40,12)
	        \fmfsurroundn{i}{4}
                 \fmfforce{(0.5w,1.7h)}{i2}
                 \fmfforce{(0.5w,-0.7h)}{i4}
               \fmflabel{$\ell$}{i1}
                 \fmflabel{$\ell$}{i3}
                 \fmf{fermion}{i3,v1}
                 \fmf{fermion}{i1,v2}
                 \fmf{phantom,left,tension=0.25,tag=1}{v1,v2}
                 \fmf{phantom,left,tension=0.25,tag=2}{v2,v1}
                 \fmfposition
                 \fmfipath{p[]}
                 \fmfiset{p1}{vpath1(__v1,__v2)}
                 \fmfiset{p2}{vpath2(__v2,__v1)}
                 \fmfi{dashes_arrow,label=$\tilde{\ell}$}{subpath (0,0.38 length(p1)) of p1}
                \fmfi{dashes_arrow,label=$\tilde{\ell}$}{subpath (length(p1), 0.63 length(p1)) of p1}
                  \fmfi{fermion,label=$\tilde{W}$}{subpath (0,length(p2)/2) of p2}
                 \fmfi{fermion,label=$\tilde{W}$}{subpath (length(p2),length(p2)/2) of p2}
                 \fmfi{wiggly}{subpath (0,length(p2)/2) of p2}
                 \fmfi{wiggly}{subpath (length(p2),length(p2)/2) of p2}
               \fmfiv{d.sh=cross,d.ang=90,d.siz=5thick}{point length(p2)/2 of p2}
\begin{fmfsubgraph}(0.25w,0.85h)(0.5w,0.5h)
	        \fmfsurroundn{j}{4}
                 \fmfforce{(0.5w,1.7h)}{j2}
                 \fmfforce{(0.5w,0.49h)}{j4}
                 \fmflabel{$h_u$}{j2}
                 \fmflabel{$h_u$}{j4} 
                 \fmf{phantom}{j3,w1}
                 \fmf{phantom}{j1,w2}
                 \fmf{phantom,left,tension=0.25,tag=1}{w1,w2}
                 \fmf{phantom,left,tension=0.25,tag=2}{w2,w1}
                 \fmfposition
                 \fmfipath{q[]}
                 \fmfiset{q1}{vpath1(__w1,__w2)}
                 \fmfiset{q2}{vpath2(__w2,__w1)}
                 \fmfi{fermion,label=$\bar{u}$}{subpath (0,length(q1))/2 of q1}
                \fmfi{fermion,label=$q$}{subpath (length(p1),length(q1)/2) of q1}
                  \fmfi{fermion,label=$\bar{u}$}{subpath (0,length(q2)/2) of q2}
                 \fmfi{fermion,label=$q$}{subpath (length(p2),length(q2)/2) of q2}
                 \fmfi{dashes_arrow}{(0.5w,1.8h) -- point length(q1)/2 of q1}
                \fmfi{dashes_arrow}{(0.5w,0.35h) -- point length(q2)/2 of q2}
 \end{fmfsubgraph} 
	        \end{fmfgraph*}
	    \end{fmffile} 

\end{center}
\vspace*{1cm}
\caption{One loop diagram and an example of a two-loop diagram contributing to neutrino masses from the dRPV $ q \bar{u} \ell^* $ operator.}
\label{Fig:neutrino}
\end{figure}
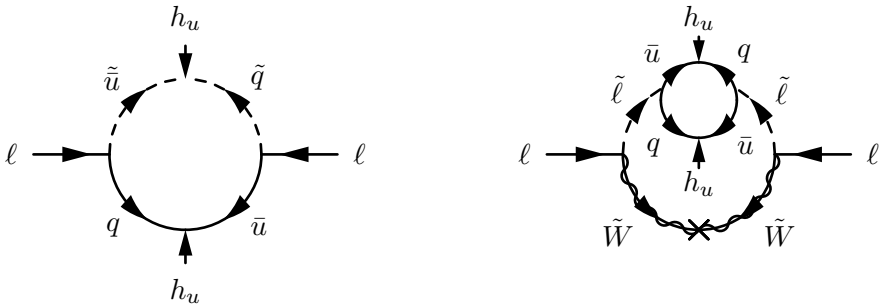
One of the strongest constraints on lepton-number violation arises from neutrino masses~\cite{Cirigliano:2005ck, Branco:2006hz}. In models of dRPV, neutrino masses can be generated at one- and two-loop from the  $ q \bar{u} \ell^* $ operator. Since Majorana neutrino masses violate lepton number by two units, two insertions of the $ q \bar{u} \ell^* $  dRPV vertex will be required to generate neutrino masses. 

The one loop diagram in Fig.~\ref{Fig:neutrino} contains the chirally suppressed part of the dRPV operators. One can estimate the magnitude of this loop-induced neutrino mass by 
\beq
m_{\nu_{ij}} \simeq \frac{1}{16\pi^2} \eta^{\prime}_{mn i}  \eta^{\prime}_{ nm j} \frac{\langle S \rangle^2 }{M^2} \frac{m^{e}_i m^{e}_j }{M^2} \frac{m^{u}_m m^{u}_n A_{u_m}}{m_{\tilde{u}_{n,m}}^2 }\ .
\eeq
Since this is proportional to quark masses as well as further flavor suppression factors from $\eta'$, we expect the largest contributions to arise from the stop/top loops, which will give  
\beq
m_{ij} \simeq 0.06~{\rm eV} \times \eta^{\prime}_{33 i}  \eta^{\prime}_{ 33 j} \left(\frac{m^{e}_i m^{e}_j}{m_\tau^2}\right) \left(\frac{S/M}{0.1} \right)^2 \left(\frac{\rm TeV}{m_{\tilde{t}_1}^2/A_t} \right) \left(\frac{10~ \rm TeV}{M}  \right)^2\,.
\eeq

The two loops contributions might be larger than the one-loop effects, since these diagrams contain the SUSY-breaking suppressed dRPV couplings rather than the chirally suppressed ones. In addition, the two-loop contributions do not require mixing in the squark sector. The magnitude of the sample two-loop diagram on the right of Fig.~\ref{Fig:neutrino} can be estimated by
\beq
m_{\nu_{ij}} \simeq \frac{g^2}{(16\pi^2)^2} \eta^{\prime \bar{\theta}^2}_{i mn} \eta^{\prime \bar{\theta}^2}_{j mn } \frac{\langle S \rangle^2 }{M^4} \frac{m^{u}_m m^{u}_n}{m_{{\rm soft}}^2} m_{\tilde{W}}\,.
\eeq
Assuming the SUSY-breaking structure of (\ref{susyepsilonx}), and again using tops in the loops, the neutrino masses are estimated to be,
\beq
m_{\nu_{ij}} \simeq 0.3~{\rm eV} \times \eta^{\prime}_{33 i}  \eta^{\prime}_{ 33 j}\left(\frac{S/M}{0.1} \right)^2 \left(\frac{\rm TeV}{m_{\tilde{\nu}_1}^2/m_{\tilde{W}}} \right) \left(\frac{\epsilon_X}{10^{-2}}  \right)^2\,.
\eeq
This can potentially be used with the appropriate FN factors to generate a realistic neutrino mass spectrum.  

\subsection{The LSP Lifetime}

The main reason behind the renewed interest in various RPV scenarios is that they might make the LSP decay inside the detector and the traditional supersymmetry searches using large missing energy will not be applicable to these models. Motivated by naturalness, it is reasonable to expect the stop to be light. Hence, here we consider the case of a stop LSP. The stop can decay via any of the dRPV operators, but we consider the case when it primarily decays via the $qq \bar{d}^*$ operator. Considering flavor suppressions, one typically expects the largest decay to be to $\tilde{t} \rightarrow \bar{b} \bar{b}$. The width is,
\beq
\Gamma_{\tilde{t}_L \rightarrow \bar{b} \bar{b}} = \frac{1}{4\pi} \left| {\eta_{333}^{\prime\prime}} \right|^2 \frac{m_b^2}{M^2} m_{\tilde{t}}  \sqrt{1- 4\frac{m_b^2}{m_{\tilde{t}}}^2 } \left(1-2 \frac{m_b^2}{m_{\tilde{t}}^2} \right)\,,
\eeq
and the lifetime is 
\begin{equation}
c \tau_{\tilde{t}} \simeq 4 {\rm~mm} \left(\frac{300  {\rm ~GeV}}{{m}_{\tilde{t}}} \right) \left(\frac{M}{ 10^8{\rm  GeV}} \right)^2
\left|\frac{1}{\eta_{333}^{\prime\prime}} \right|^2 \,.
\end{equation}
While it is likely at least somewhat displaced, for reasonable choices of the parameters the decay happens inside the detector.

\section{Froggatt-Nielsen Suppression Factors\label{App:FNsuppression}}

Here we consider a specific set of Froggatt-Nielsen charges and  flavor suppression factors of the coefficients $\eta, \eta', \eta''$ defined in Eq.~\eqref{eq:KRPV}, as well as the coefficient $\lambda^{\not{B},\not{L}}$ of the R-parity invariant B and L violating, dimension 6 operator presented in Eq.~\eqref{Kdrpv2}. The lepton and quark charges are taken from \cite{Leurer:1992wg} and \cite{Leurer:1993gy}, respectively.  In this case there are two separate $U(1)_{\rm FN}$ horizontal symmetries, corresponding to two spurions which introduce the suppressions, 
\begin{equation}
\epsilon_1 \sim \epsilon \sim 0.2, \ \ \epsilon_2 \sim \epsilon^2\sim 0.04\,.
\end{equation}
The overall suppression factor will be the product from the two symmetries. The flavor suppression factors from this symmetry are given in the left side of Tables~\ref{tab:etaetapFN}~and~\ref{tab:lambdaFN}. These are sufficient to evade all bounds from baryon and lepton number violation even for $\epsilon \sim 0.1$. At the same time, the largest of them are sufficient to make a third generation LSP decay inside the collider. 

Since $\eta''$ is symmetric in the $q$'s, it has 18 independent components and we present the suppression factors for all of them in the bottom panel of Table~\ref{tab:etaetapFN}. However $\eta$ and $\eta'$ have 27 components, so for these we only present those that are relevant for proton decay in the top and middle panels of Table~\ref{tab:etaetapFN}. Finally,  $\lambda^{\not{B},\not{L}}$ has 54 independent components and we present only those that can contribute to the proton decay amplitude in Table~\ref{tab:lambdaFN}.

\section{Useful Supersymmetric Identities\label{App:SUSYidentities}}

All superspace conventions follow Martin's Supersymmetric Primer \cite{Martin:1997ns}. The Lagrangian from a generic non-holomorphic dPRV operator is
\begin{eqnarray}
&&\int d^4 \theta ({\bf a}+{\bf b}\theta^2+{\bf c}\bar{\theta}^2 + {\bf d} \theta^2 \bar{\theta}^2 )\Phi_j \Phi_k \Phi^{*i} \\
&&={\bf a} \big( i (\phi_j \psi_k + \psi_j \phi_k )\sigma^\mu \partial_\mu \psi^{\dagger i} -  \psi_j \psi_k F^{*i} + \phi_j \phi_k \partial_\mu \partial^\mu \phi^{*i}+ (\phi_k F_j + \phi_j F_k  ) F^{*i} \big) \no \\
&&~~ + {\bf b} \left( \phi_j \phi_k F^{* i} \right) +   {\bf c}\left(\psi_j \psi_k \phi^{*i} -\phi_j F_k \phi^{*i}  + \phi_k F_j \phi^{*i} \right)+ {\bf d} \left( \phi_j \phi_k \phi^{*i} \right).\no
\label{kahlerexpand}
\end{eqnarray}
The supersymmetric tree-level equations of motion, with a canonical Kahler potential is given by,
\beq
\frac{1}{4} \mathcal{D}^2 \Phi= \frac{\delta W^*}{\delta \Phi^{*} } ,\label{eq:EOM}
\eeq
where in components, the left hand side is,
\beq
- \frac{1}{4} \mathcal{D}^{ 2}  \Phi(y) =  F(y) - i \sqrt{2} \bar{\theta} \bar{\sigma}^\mu \partial_\mu \psi(y)  +\bar{\theta}^2 \partial_\mu \partial^\mu\phi(y)\,,
\label{eq:Dsquare}
\eeq
and $ y^\mu = x^\mu + i \bar{\theta} \bar{\sigma}^\mu \theta$.
A D-term can be rewritten as an F-term using the identity,
\beq
\int d^4 \theta V = \int d^2 \theta \left(-\frac{1}{4} \mathcal{D}^{\dagger 2} V \right).
\label{eq:DtoF}
\eeq

\begin{table}[h!]
\begin{center}
\begin{tabular}{c|ccc|ccc|ccc|ccc|ccc}
 &$q_1$ & $q_2$ & $q_3$ & $\bar{u}_1$ & $\bar{u}_2$ &$ \bar{u}_3$ & $\bar{d}_1$ & $\bar{d}_2$ & $\bar{d}_3$ & $L_1$ & $L_2$ & $L_3$ & $\bar{e}_1$ & $\bar{e}_2$ & $\bar{e}_3$ \\ \hline\hline
 $U(1)_1$ & -5 & -2 & 0 & 11 & 3 & 0 & 7 & 6 & 2 & 1 & 1 & 0 & 0 & 0 & 0 \\
  $U(1)_2$ & 4 & 2 & 0 & -4 & -1 & 0 & -2 & -2 & 0 & 1 & 0 & 0 & 1 & 1 & 1 \\
\end{tabular}
\caption{FN charges for a model with two horizontal $U(1)$ symmetries and flavor breaking spurions. The lepton and quark charges are taken from \cite{Leurer:1992wg} and \cite{Leurer:1993gy}, respectively.  \label{tab:FNcharges}}
\end{center}
\end{table}
\begin{table}[h!]
\begin{center}
\[\arraycolsep=2pt\def\arraystretch{1.2} \begin{array}{ccc}

\begin{array}{c|cccc}
\eta  & \bar{u}_1\bar{e}_1 & \bar{u}_1\bar{e}_2 & \bar{u}_2\bar{e}_1 &
   \bar{u}_2\bar{e}_2 \\ \hline
 \bar{d}_1 & \epsilon ^6 & \epsilon ^6 & \epsilon ^8 & \epsilon ^8 \\
 \bar{d}_2 & \epsilon ^7 & \epsilon ^7 & \epsilon ^7 & \epsilon ^7 \\
 \bar{d}_3 & \epsilon ^{15} & \epsilon ^{15} & \epsilon  & \epsilon  \\
\end{array} &~~~~~&
\begin{array}{c|cccc}
 \eta & \bar{u}_1\bar{e}_1 & \bar{u}_1\bar{e}_2 & \bar{u}_2\bar{e}_1 &
   \bar{u}_2\bar{e}_2 \\ \hline
 \bar{d}_1 & \epsilon ^{28} & \epsilon ^{28} & \epsilon ^{14} & \epsilon
   ^{14} \\
 \bar{d}_2 & \epsilon ^{27} & \epsilon ^{27} & \epsilon ^{13} & \epsilon
   ^{13} \\
 \bar{d}_3 & \epsilon ^{19} & \epsilon ^{19} & \epsilon ^5 & \epsilon ^5 \\
\end{array}\\
~\\
\begin{array}{c|cccc}
 \eta' & \bar{u}_1\ell _1^* & \bar{u}_1\ell _2^* & \bar{u}_2\ell _1^* &
   \bar{u}_2\ell _2^* \\ \hline
 q_1 & \epsilon ^7 & \epsilon ^5 & \epsilon ^7 & \epsilon ^9 \\
 q_2 & \epsilon ^{14} & \epsilon ^{12} & 1 & \epsilon ^2 \\
 q_3 & \epsilon ^{20} & \epsilon ^{18} & \epsilon ^6 & \epsilon ^4 \\
\end{array} &~~~~~&
\begin{array}{c|cccc}
  \eta' & \bar{u}_1\ell _1^* & \bar{u}_1\ell _2^* & \bar{u}_2\ell _1^* &
   \bar{u}_2\ell _2^* \\ \hline
 q_1 & \epsilon ^9 & \epsilon ^7 & \epsilon ^{11} & \epsilon ^9 \\
 q_2 & \epsilon ^{16} & \epsilon ^{14} & \epsilon ^6 & \epsilon ^4 \\
 q_3 & \epsilon ^{22} & \epsilon ^{20} & \epsilon ^8 & \epsilon ^6 \\
\end{array}\\
~\\
\begin{array}{c|cccccc}
 \eta'' & q_1q_1 & q_1q_2 & q_2q_2 & q_1q_3 & q_2q_3 & q_3q_3 \\ \hline
 \bar{d}_1 & \epsilon ^{37} & \epsilon ^{30} & \epsilon ^{23} & \epsilon
   ^{24} & \epsilon ^{17} & \epsilon ^{11} \\
 \bar{d}_2 & \epsilon ^{36} & \epsilon ^{29} & \epsilon ^{22} & \epsilon
   ^{23} & \epsilon ^{16} & \epsilon ^{10} \\
 \bar{d}_3 & \epsilon ^{28} & \epsilon ^{21} & \epsilon ^{14} & \epsilon
   ^{15} & \epsilon ^8 & \epsilon ^2 \\
\end{array} &~~~~~&
\begin{array}{c|cccccc}
\eta''  & q_1q_1 & q_1q_2 & q_2q_2 & q_1q_3 & q_2q_3 & q_3q_3 \\ \hline
 \bar{d}_1 & \epsilon ^{37} & \epsilon ^{30} & \epsilon ^{23} & \epsilon
   ^{32} & \epsilon ^{17} & \epsilon ^{11} \\
 \bar{d}_2 & \epsilon ^{36} & \epsilon ^{29} & \epsilon ^{22} & \epsilon
   ^{24} & \epsilon ^{16} & \epsilon ^{10} \\
 \bar{d}_3 & \epsilon ^{28} & \epsilon ^{21} & \epsilon ^{14} & \epsilon
   ^{15} & \epsilon ^8 & \epsilon ^2 \\
\end{array}\\
\end{array}\]
\caption{FN suppression of $\eta_{ijk}, \eta^{\prime}_{ijk}, \eta^{\prime\prime}_{ijk}$ for the charges given in Table \ref{tab:FNcharges} applied to the case of flavor breaking external to the dRPV sector ({\bf left}) and   for the case of flavor breaking and dRPV mediation sectors unified ({\bf right}). \label{tab:etaetapFN}}
\end{center}
\end{table}
\begin{table}[h!]
\begin{center}
\[\arraycolsep=2pt\def\arraystretch{1.2} \begin{array}{ccc}
\begin{array}{c|cccc}
 \lambda^{\not{B},\not{L}} & \bar{u}_1\bar{e}_1 & \bar{u}_1\bar{e}_2 & \bar{u}_2\bar{e}_1 &
   \bar{u}_2\bar{e}_2 \\ \hline
 q_1q_1 & \epsilon ^{43} & \epsilon ^{43} & \epsilon ^{29} & \epsilon ^{29}
   \\
 q_1q_2 & \epsilon ^{36} & \epsilon ^{36} & \epsilon ^{22} & \epsilon ^{22}
   \\
 q_2q_2 & \epsilon ^{29} & \epsilon ^{29} & \epsilon ^{15} & \epsilon ^{15}
   \\
\end{array} &~~~~~&
\begin{array}{c|cccc}
\lambda^{\not{B},\not{L}}  & \bar{u}_1\bar{e}_1 & \bar{u}_1\bar{e}_2 & \bar{u}_2\bar{e}_1 &
   \bar{u}_2\bar{e}_2 \\ \hline
 q_1q_1 & \epsilon ^{23} & \epsilon ^{23} & \epsilon ^{23} & \epsilon ^{23}
   \\
 q_1q_2 & \epsilon ^{22} & \epsilon ^{22} & \epsilon ^{16} & \epsilon ^{16}
   \\
 q_2q_2 & \epsilon ^{21} & \epsilon ^{21} & \epsilon ^9 & \epsilon ^9 \\
\end{array}

\end{array}\]
\caption{FN suppression of $\lambda^{\not{B},\not{L}}_{ijmn}$ for the charges given in Table \ref{tab:FNcharges} applied to the case of flavor breaking external to the dRPV sector ({\bf left}) and   for the case of flavor breaking and dRPV mediation sectors unified ({\bf right}). \label{tab:lambdaFN}}
\end{center}
\end{table}
\newpage

\bibliographystyle{ieeetr}

\end{document}